\documentclass[fleqn,usenatbib]{mnras}
\pdfoutput=1

\usepackage[T1]{fontenc}

\DeclareRobustCommand{\VAN}[3]{#2}
\let\VANthebibliography\thebibliography
\def\thebibliography{\DeclareRobustCommand{\VAN}[3]{##3}\VANthebibliography}

\usepackage{graphicx}	
\usepackage{amsmath}	
\usepackage{amssymb}	
\usepackage{latexsym}
\usepackage{xcolor}
\usepackage[utf8]{inputenc}
\usepackage[utf8]{luainputenc}
\usepackage{babel}
\usepackage{float}

\usepackage{bm}

\usepackage{import}
\usepackage{siunitx}
\DeclareSIUnit \parsec {pc}
\DeclareSIUnit \year {yr}
\usepackage[plain]{algorithm}
\usepackage{caption}
\usepackage[noend]{algpseudocode}
\definecolor{newcolor}{rgb}{.8,.349,.1}

\newcommand{\bb}[1]{\left(#1\right)}                                        
\newcommand{\bs}[1]{\left\{#1\right\}}                                      
\newcommand{\abs}[1]{\left|#1\right|}                                       

\newcommand{\eref}[1]{Eq.~\ref{eq:#1}}

\newcommand{\fref}[1]{Fig.~\ref{fig:#1}}
\newcommand{\figref}[1]{Figure~\ref{fig:#1}}
\newcommand{\secref}[1]{Section~\ref{sec:#1}}
\newcommand{\sref}[1]{\ref{sec:#1}}
\newcommand{\stepref}[1]{Step~\ref{step:#1}}

\newcommand{\knusbar}[0]{\bar{k}_{\nu,\mathrm{s}}}
\newcommand{\knus}[0]{k_{\nu,\mathrm{s}}}
\newcommand{\knuabar}[0]{\bar{k}_{\nu,\mathrm{a}}}

\newcommand{\Nrot}[0]{N_{\mathrm{rot}}}

\newcommand{\trec}[0]{t_{\mathrm{rec}}}
\newcommand{\direc}[0]{\bm{\Omega}}
\newcommand{\nelectron}[0]{n_{\mathrm{e}}}
\newcommand{\nH}[0]{n_{\mathrm{H}}}
\newcommand{\alphaB}[0]{\alpha_{\mathrm{B}}}
\newcommand{\Rstroemgren}[0]{R_{\mathrm{St}}}
\newcommand{\cs}[0]{c_{\mathrm{s}}}
\newcommand{\Rr}[0]{R_{\mathrm{r}}}
\newcommand{\Rd}[0]{R_{\mathrm{d}}}

\newcommand{\xHP}[0]{x_{\mathrm{H+}}}
\newcommand{\ndim}[0]{n_{\mathrm{dim}}}

\newcommand{\nItPbc}[0]{n_{\mathrm{it, pbc}}}
\newcommand{\nItScat}[0]{n_{\mathrm{it, scat}}}

\newcommand{\nbase}[0]{n_{\mathrm{base}}}
\newcommand{\tbase}[0]{t_{\mathrm{base}}}
\newcommand{\ttask}[0]{t_{\mathrm{task}}}
\newcommand{\tideal}[0]{t_{\mathrm{ideal}}}
\newcommand{\Sideal}[0]{S_{\mathrm{ideal}}}
\newcommand{\sigmaS}[0]{\sigma_{\mathrm{s}}}

\title[Radiative transfer via transport sweeps]{The Sweep Method for radiative Transfer in Arepo}

\author[Peter et al.]{
Toni Peter,$^{1}$\thanks{E-mail: toni.peter@uni-heidelberg.de}
Ralf S. Klessen,$^{1}$
Guido Kanschat,$^{2}$
Simon C. O. Glover,$^{1}$
Peter Bastian$^{2}$
\\
$^{1}$Universit\"{a}t Heidelberg, Zentrum f\"{u}r Astronomie, Institut f\"{u}r Theoretische Astrophysik, Albert-Ueberle-Str.\ 2, 69120 Heidelberg, Germany\\
$^{2}$Universit\"{a}t Heidelberg, Interdisziplin\"{a}res Zentrum f\"{u}r Wissenschaftliches Rechnen, Im Neuenheimer Feld 225, 69120 Heidelberg, Germany \\
}

\date{Accepted XXX. Received YYY; in original form ZZZ}

\pubyear{2022}

\begin{document}
\label{firstpage}
\pagerange{\pageref{firstpage}--\pageref{lastpage}}
\maketitle

\begin{abstract}
We introduce the radiative transfer code Sweep for the cosmological simulation suite Arepo. Sweep is a discrete ordinates method in which the radiative transfer equation is solved under the infinite speed of light, steady state assumption by a transport sweep across the entire computational grid. Since Arepo is based on an adaptive, unstructured grid, the dependency graph induced by the sweep dependencies of the grid cells is non-trivial. In order to solve the topological sorting problem in a distributed manner, we employ a task-based-parallelism approach. The main advantage of the sweep method is that the computational cost scales only with the size of the grid, and is independent of the number of sources or the distribution of sources in the computational domain, which is an advantage for radiative transfer in cosmological simulations, where there are large numbers of sparsely distributed sources. We successfully apply the code to a number of physical tests such as the expansion of HII regions, the formation of shadows behind dense objects, the scattering of light, as well as its behavior in the presence of periodic boundary conditions. In addition, we measure its computational performance with a focus on highly parallel, large-scale simulations.
\end{abstract}

\begin{keywords}
cosmology -- radiative transfer -- radiation dynamics -- ISM: HII regions
\end{keywords}



\section{Introduction}
The era of reionization is an important period in the history of the universe, during which the composition of the intergalactic medium transitioned from mostly neutral to highly ionized. This period marks an important transition between the early universe which was largely homogeneous with small fluctuations and the highly structured and complex universe we see at present days \citep[see e.g.][]{Zaroubi13,wise19,loebReionizationUniverseFirst2001}.

One way to understand the process of reionization is with numerical simulations.
However, modeling reionization is a numerically challenging problem.
Whereas the physics of the early universe was dominated by gravity, reionization is driven by the first stars and galaxies.
In order to understand reionization, it is necessary to accurately model the formation and feedback processes of these small objects.
The small dwarf galaxies which are believed to be the dominant sources of ionizing photons only have sizes of $\sim \SI{1}{\kilo\parsec}$ in size, whereas in order to obtain representative samples, the simulated volume of space needs to be sufficiently large, with lengths on the order of hundreds of Mpc \citep{iliev14}.
This implies a vast range of length scales that need to be represented in any numerical model. The need to simulate such large volumes of space also implies that we must be able to follow the effects of ionizing radiation from a very large number of sources. Together, these requirements strongly constrain our choice of algorithm for modeling the transport of ionizing photons in the early Universe. For example, ray tracing with long characteristics \citep{mihalasFoundationsRadiationHydrodynamics1999, abelPhotonconservingRadiativeTransfer1999, whalenMultistepAlgorithmRadiation2006}, a method which has been used with great success to model individual HII regions in the local Universe \citep[e.g.][]{peters10,kim18} is completely impractical in this context, as its computational cost scales as the product of the number of ionizing sources and the number of resolution elements in the simulation, $N_{\rm source} 
\times N_{\rm cell}$.
This motivates the search for approaches that are independent of the number of ionizing sources.

In this paper, we focus on the astrophysical simulation package {\sc Arepo}~\citep{springelPurSiMuove2010}. {\sc Arepo} solves the gravitational equation and the magnetohydrodynamical equations for a magnetized gas
on a co-moving Voronoi grid. It also has different physics modules, including treatments of stellar feedback (supernovae, radiation) and non-equilibrium chemistry. The main goal of this project is optimizing the performance of radiative transfer in {\sc arepo}.

Radiative transfer is an especially challenging problem for numerical simulations for a number of reasons~\citep{mihalasFoundationsRadiationHydrodynamics1999}.
The first is the high dimensionality of the relevant physical quantity: radiation intensity, which depends on three spatial, two directional, one temporal and one frequency dimension leading to a total of seven dimensions.
Furthermore, the properties of the local medium, such as the emissivity, absorptivity and fraction of scattered photons are important for the solution of the radiative equation while at the same time being dependent on the radiation, thus creating a need for iterative schemes to obtain solutions of the full equations.
In addition, the radiative transfer equation changes its mathematical properties from being elliptical in optically thick regions to being hyperbolic in optically thin regions, thus making it difficult to choose a specialized solver suited for a particular type of equation that works across all scales of optical depth.

One class of methods with this property are moment-based methods, where one solves the moments of the radiative transfer equation with some approximate closure relation.
This can lead to drastically improved performance at the cost of precision.
A number of moment-based methods exists, which differ primarily in the choice of closure relation, which is typically given by an approximate expression for the Eddington tensor.
One example of a moment-based method is the flux limited diffusion approach~\citep{levermoreFluxlimitedDiffusionTheory1981, whitehouseSmoothedParticleHydrodynamics2004} in which the closure relation is derived under the assumption of slowly varying intensity and the purpose of the flux limiter is to ensure that changes in the radiation field cannot propagate faster than the speed of light. Flux-limited diffusion has been applied to various astrophysical problems \citep[e.g.][]{Krumholz2007,Boss2008}, but the high diffiusivity of the method and its consequent inability to properly account for shadowing \citep[see e.g.][]{Hayes2003} make it a poor choice for modelling ionizing radiation.

A moment-based method with a different closure relation is given by the optically thin variable Eddington tensor method in which the Eddington tensor is calculated by assuming that all lines of sight to the sources in the simulation are optically thin \citep{Gnedin01}. This algorithm is efficient, but its accuracy is highly problem-dependent.

The radiative transfer equation can also be solved by Monte Carlo methods, in which rays are represented by photon packets~\citep{oxleySmoothedParticleHydrodynamics2003,dullemondRADMC3DMultipurposeRadiative2012}. Each photon packet is appropriately sampled from the distribution of sources which then interact with the gas according to their properties. This approach has the advantage of requiring few approximations to the equations themselves, so that the quality of the results is determined primarily by the number of photon packets emitted. A disadvantage of this approach is the presence of statistical noise, with a signal to noise ratio that scales as $\mathrm{SNR} \propto \sqrt{n}$, where $n$ is the number of photon packets. In addition, this method is difficult to parallelize in situations where duplicating the entire grid structure on every processor is impractical owing to the memory requirements, a situation we often find ourselves in when simulating e.g.\ cosmic reionization.

In this paper, we focus on the astrophysical simulation package {\sc Arepo}~\citep{springelPurSiMuove2010}. {\sc Arepo} solves the gravitational equation and the magnetohydrodynamical equations for a magnetized gas
on a co-moving Voronoi grid. It also has different physics modules, including treatments of stellar feedback (supernovae, radiation) and non-equilibrium chemistry. The main goal of this project is optimizing the performance of radiative transfer in {\sc arepo}.

Some of the currently available methods for radiative transfer in {\sc Arepo} that have computational costs that are largely independent of the number of sources are the M1 method, which is a moment-based method based on the M1 closure relation~\citep{kannanArepoRTRadiationHydrodynamics2019}, the Monte-Carlo radiation hydrodynamics method MCRT~\citep{smithAREPOMCRTMonteCarlo2020} and the SimpleX method~\citep{ritzerveldTransportAdaptiveRandom2006,jiangAlgorithmRadiationMagnetohydrodynamics2014,changTimedependentRadiationHydrodynamics2020}.
While the M1 method is comparatively fast, it suffers from numerical problems inherent to moment-based methods, such as the two-beam instability~\citep{rosdahlRamsesrtRadiationHydrodynamics2013}.
The MCRT method employs a number of techniques to improve upon Monte-Carlo radiative transfer. Currently, it is not viable to perform simulations of galaxy formation with this approach but improvements to the method are still in active development.
The original SimpleX method is similar to a short-characteristics scheme and does not require angular discretization.
However, it suffers from numerical diffusion, which was the reason for the development of SimpleX2~\citep{paardekooperSimpleX2RadiativeTransfer2010} and its implementation in {\sc Arepo}, SPRAI~\citep{jauraSPRAICouplingRadiative2018,jauraSPRAIIIMultifrequencyRadiative2020}.
In these methods, angular discretization is introduced, effectively making them discrete ordinate methods.
Discrete ordinates methods have the advantage that they do not require any physically motivated approximations such that in principle, any numerical artefacts can be reduced by an increase in the resolution.

Simplex2 and SPRAI work as follows.
At the beginning of every time step, photons are created at source cells and distributed equally into all directional bins.
Then, in every iteration, photons from a directional bin are transported from a Voronoi cell to its $d$ most straightforward neighbors along that direction.
The photon density is then used to update the local chemistry of the cell and some of the photons are scattered by re-distributing into the other directional bins.
This process is iterated until all of the photons have been absorbed.
This method performs well in optically thick regions in which the mean free path is short.
However, in optically thin regions, this method requires many iterations, increasing computation times drastically.

Our proposed change to this algorithm is based on transport sweeps~\citep{kochSolutionFirstorderForm1991}. The idea is that, for a given direction, a cell is only solved once all its upwind neighbors along that direction have been solved. The main benefit of this method is that, in the absence of scattering, such a re-ordering allows us to obtain the full photon density field in a single sweep. In order to incorporate scattering, the sweep needs to be iterated.

The drawback of this method is that it induces an ordering on the cells due to the dependencies of cells on their upwind neighbors. While the dependency graph is trivial for regular grids, this is not the case for a Voronoi grid. At the same time, the code needs to be parallelized. Our current solution to these problems is task-based parallelism~\citep{zeyaoParallelFluxSweep2004} in which a task is a pair of a Voronoi cell and a given directional bin. For each task, we keep track of the number of unsolved upwind neighbors and only solve those tasks for which this number is zero. In this way, the dependency graph is never explicitly constructed but we still obtain a topological ordering of the cells.

This paper is structured as follows.
In \secref{methods} we discuss the problem of radiative transfer in general (\secref{radiativeTransfer}) before the concept of radiative transport sweeps (\secref{sweep}) and the concrete implementation details which allow the code to run on large numbers of processors in parallel (\secref{parallel}) are introduced. We also discuss how to handle problems with periodic boundaries in the concept of transport sweeps (\secref{methodspbc}).
In \secref{tests}, we present a number of tests in order to demonstrate that our code reproduces physically correct results (\secref{expansionTests}, \sref{shadowing}, \sref{scattering}, \sref{pbcTest}). We also study the computational performance of the code, especially in respect to its parallelization in \secref{strongScaling} and \sref{weakScaling}.
Finally, we conclude this paper and present some potential extensions of the code as well as possible applications in \secref{conclusion}.

\section{Methods}\label{sec:methods}
\subsection{Structure of the code}
In this paper we discuss an implementation of radiative transfer for the astrophysical simulation code {\sc Arepo} \citep{springelPurSiMuove2010}.
The structure of our code is based very closely on SPRAI, an existing radiative transfer module for {\sc Arepo} whose design and operation is described in \citet{jauraSPRAICouplingRadiative2018,jauraSPRAIIIMultifrequencyRadiative2020}. Indeed, the code shares SPRAI's interface between radiative transfer and the SGChem chemistry module.\footnote{SGChem implements various different chemical networks. In this paper, we use its primordial chemistry network, first implemented in {\sc Arepo} by \citet{hartwigHowImprovedImplementation2015} and more recently updated by \citet{schauerInfluenceStreamingVelocities2019}} We therefore do not discuss this aspect of the code here and refer the reader interested in details of this coupling to \citet{jauraSPRAICouplingRadiative2018,jauraSPRAIIIMultifrequencyRadiative2020}.

The defining characteristic of {\sc Arepo} is that the hydrodynamical equations are solved on a Voronoi grid which is generated by points that are co-moving with the gas instead of a Eulerian grid with an adaptive mesh refinement scheme.
This has the benefit of avoiding numerical artifacts caused by the structure of the grid while simultaneously adapting the grid to the gas density automatically.

The goal of the radiative transfer code is to solve the radiative transfer equation to obtain the radiative fluxes in all cells.
These fluxes are then passed to the chemistry module which requires the fluxes to calculate the detailed chemical composition of the medium as well as the corresponding heating and ionization rates.

{\sc Arepo} uses an adaptive timestepping approach in which regions that require higher accuracy are solved with a smaller timestep.
Due to this, the full Voronoi grid is only available on the synchronization timesteps, i.e.\ the timesteps during which every cell is updated.
Our current implementation of the radiative transfer method requires the full grid, so that we can only perform radiative transfer calculations during those synchronization steps.
For the substeps, the radiative fluxes between cells are assumed to remain constant, keeping the value of the previous synchronization step.
The validity of this approximation depends on the ratio of the lowest hydrodynamical/gravity timestep to the full synchronization timestep as well as on the ratio of the timescale at which hydrodynamics and gravity take place to the timescale of radiative transfer and the photochemistry.
At the cost of code performance, the effect of the approximation can be reduced by limiting the highest timestep.
For the tests performed in this paper, this approximation has been acceptable.
In the future, extensions to the implementaton can be considered in which radiative transfer takes place on the substeps as well, which would require adjusting the cell timestep criterion to take radiative transfer into account.

\subsection{Radiative transfer\label{sec:radiativeTransfer}}
The quantity of interest in the problem of radiative transfer is the specific radiative intensity $I_{\nu}(\bm{r}, t, \hat{\Omega})$, with frequency $\nu$, spatial position $\bm{r}$, time $t$ and solid angle $\hat{\Omega}$ given in units of $\SI{}{\watt\per\meter\squared\per\steradian\per\hertz}$. 
The radiative transfer equation is given by~\citep{rybickiRadiativeProcessesAstrophysics1985}.
\begin{align}
\frac{1}{c}\frac{\partial}{\partial t}I_\nu + \hat{\bm{\Omega}} \cdot \bm{\nabla} I_\nu = j_\nu - (\knusbar+\knuabar) I_\nu + \frac{1}{4\pi} \int_{S} \knus(\bm{\Omega}') I_\nu \bm{\mathrm{d}\Omega'}. 
\label{eq:radiativeTransferGeneral}
\end{align}
It relates the rate of change of the radiative intensity along the line with solid angle $\hat{\bm{\Omega}}$. Here, $c$ is the speed of light, $j_{\nu}$ is the emission coefficient, $\bar{k}_{\nu, \mathrm{a}}$ is the absorption coefficient, $\int_{S}$ denotes the integral over the unit sphere with $\bm{\Omega'}$ being the solid angle relative to $\hat{\bm{\Omega}}$.
The total scattering coefficient $\knusbar$ is defined via the angle-dependent scattering coefficient $\knus(\bm{\Omega'})$ as $\knusbar = \int_{S} \knus(\bm{\Omega'}) \bm{\mathrm{d}\Omega'}$.
From now on, we will assume isotropic scattering such that $\knus(\bm{\Omega}) = \knusbar$.
If the timescales on which the material coefficients (the source term and the absorption and scattering coefficients) change are all smaller than the typical light crossing time of the system, we can safely make the so-called infinite speed of light approximation in which we drop the first term of \eref{radiativeTransferGeneral}. 

With these two assumptions we obtain
\begin{align}
\hat{\bm{\Omega}} \cdot \bm{\nabla} I_\nu = j_\nu - (\knusbar + \knuabar) I_\nu + \frac{\knusbar}{4\pi} \int_{S}  I_\nu \bm{\mathrm{d}\Omega'}. 
\label{eq:radiativeTransfer}
\end{align}
In order to solve this equation numerically, we need to find a discretization scheme, of which there are many for the radiative transfer problem.
Here, we will focus on the discrete ordinate method in which the equation is discretized in all variables: time $t$, position $\bm{r}$, frequency, $\nu$ and the angular component $\Omega$.
\begin{figure}
    \centering
    \def\svgwidth{\columnwidth}
    \fontsize{13pt}{31pt}\selectfont
\begingroup%
  \makeatletter%
  \providecommand\color[2][]{%
    \errmessage{(Inkscape) Color is used for the text in Inkscape, but the package 'color.sty' is not loaded}%
    \renewcommand\color[2][]{}%
  }%
  \providecommand\transparent[1]{%
    \errmessage{(Inkscape) Transparency is used (non-zero) for the text in Inkscape, but the package 'transparent.sty' is not loaded}%
    \renewcommand\transparent[1]{}%
  }%
  \providecommand\rotatebox[2]{#2}%
  \newcommand*\fsize{\dimexpr\f@size pt\relax}%
  \newcommand*\lineheight[1]{\fontsize{\fsize}{#1\fsize}\selectfont}%
  \ifx\svgwidth\undefined%
    \setlength{\unitlength}{96.73463612bp}%
    \ifx\svgscale\undefined%
      \relax%
    \else%
      \setlength{\unitlength}{\unitlength * \real{\svgscale}}%
    \fi%
  \else%
    \setlength{\unitlength}{\svgwidth}%
  \fi%
  \global\let\svgwidth\undefined%
  \global\let\svgscale\undefined%
  \makeatother%
  \begin{picture}(1,0.85727904)%
    \lineheight{1}%
    \setlength\tabcolsep{0pt}%
    \put(0,0){\includegraphics[width=\unitlength,page=1]{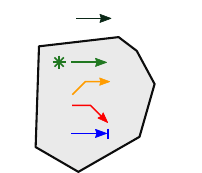}}%
    \put(0.42210945,0.79947475){\color[rgb]{0,0,0}\makebox(0,0)[lt]{\lineheight{1.25}\smash{\begin{tabular}[t]{l}$\bm{\Omega}$\end{tabular}}}}%
    \put(0,0){\includegraphics[width=\unitlength,page=2]{radiativeProcesses.pdf}}%
  \end{picture}%
\endgroup%

    \caption{Illustration of the radiative processes described by \eref{radiativeTransfer} for a single grid cell: Incoming (brown) and outgoing (teal) radiation, sources (green), absorption (blue) and scattering (into considered solid angle: orange, out of it: red)\label{fig:radiativeProcesses}}
\end{figure}
The physical intuition behind \eref{radiativeTransfer} when applied to a small volume is illustrated in \fref{radiativeProcesses}.
The sources of radiation in this cell are through incoming radiation from cells to the left (brown arrows), the source term $j$ directly (green arrow) or scattering into $\direc$ from a different $\direc^{\prime}$ (orange arrow). Radiation from the cell is either scattered out of this $\direc$ (red arrow), absorbed (blue arrow) or leaves the cell towards the right (teal arrows).
Thus, the neighboring cells fall into two categories:
Cells upwind along $\direc$ (brown arrows) -- in order to solve the local equations, we require the incoming specific intensities from those cells.
Cells downwind along $\direc$ (teal arrows) -- these depend on the local solution of the intensity for their own solution.

The discretized radiative transfer equation takes the form of a large, coupled system of equations.
There are many different approaches to solving this problem.
Which of these methods is the best strongly depends on the physical nature of the simulation.
In very optically thick media, scattering dominates, which means the equations are elliptical and thus diffusive in nature.
On the other hand, in optically thin media, the equations become hyperbolic and long-ranged.

\subsection{Source iteration\label{sec:sourceIteration}}
An obvious approach to solve the resulting equations is to construct the full matrix describing the system and to apply an iterative solver such as the Generalized minimal residual method~\citep{saadGMRESGeneralizedMinimal1986} until convergence is reached. Due to the high dimensionality of the equation (three spatial, two angular and one frequency dimension), this quickly becomes infeasible due to the sheer size of the resulting matrix.

A different, well-known approach is known as source iteration which is given by Algorithm~\ref{alg:sourceIteration}.
Here, convergence of $I_{\nu}$ can be defined in a number of ways. The definition we choose is given by the condition 
\begin{align}
    \forall \bm{r} \forall \bm{\Omega}: \frac{\abs{I_{\nu}^i(\bm{r}, \bm{\Omega}) - I_{\nu}^{i-1}(\bm{r}, \bm{\Omega})}}{I_{\nu}^{i-1}(\bm{r}, \bm{\Omega})} < \epsilon,
\end{align}
where $\epsilon$ is a free parameter and should be chosen to be small.

\begin{algorithm}
\caption{Source iteration}\label{alg:sourceIteration}
\begin{algorithmic}[1]
\State Guess initial intensity $I_{\nu}^0$. For example: $I_{\nu}^0 = 0$.
\While{$I_{\nu}^i$ not converged}
\State Compute source terms (using $I_{\nu}^i$ for scattering).
\State Solve \eref{radiativeTransfer} to obtain $I_{\nu}^{i+1}$ using source terms for each $\Omega$ and
\par each $\nu$.\label{step:sweep}
\EndWhile
\end{algorithmic}
\end{algorithm}

The idea is to use an iterative scheme in which scattering is treated as a constant source term.
This is still an iterative method, as the scattering source terms are re-computed after every iteration.
This approach is suited best for optically thin media where scattering is not dominant and the source iteration converges quickly.

The main benefit of this is that it removes the coupling between the terms of different $\Omega$, so that instead of solving one large coupled system of equations, \stepref{sweep} of Algorithm~\ref{alg:sourceIteration} only requires us to solve one smaller system of equations for each $\Omega$.
In the following, we will show that there is an efficient way to solve such a system of equations under certain conditions.
In principle, we would like to simply iterate over the grid once and solve \eref{radiativeTransfer} for each grid cell to obtain an exact solution.
However, as discussed previously, the equation gives rise to local dependencies that require us to solve upwind cells before their downwind neighbors.

\begin{figure}
    \centering
    \def\svgwidth{\columnwidth}
    \fontsize{13pt}{31pt}\selectfont
    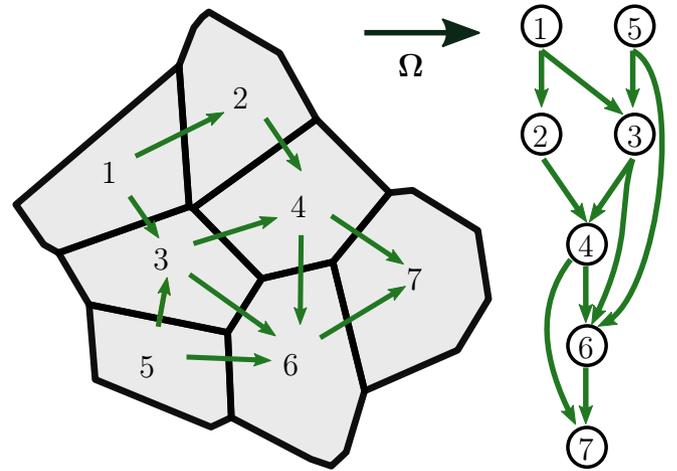
    \caption{\label{fig:gridCells}Left: Illustration of a 2D Voronoi grid and the dependencies induced by the sweep ordering for a sweep towards the right. Right: The directed, acylic graph corresponding to the dependencies.}
\end{figure}

As illustrated in \fref{gridCells}, we can understand these local dependencies as a directed graph in which the nodes are the grid cells and an edge from the cell $c_1$ to the cell $c_2$ corresponds to a dependency of $c_2$ on $c_1$.
Under the assumption that the graph is acyclic (which we can easily prove to be true for the induced dependency graph of a Voronoi grid, such as the one used in {\sc Arepo}; see Appendix~\ref{sec:voronoiCycleProof}), there is a topological ordering of the grid cells, such that any cell in the ordering only depends on the cells that come before it.
This is analogous to a re-ordering of the cells of the matrix describing the system of equations (in which each non-zero entry corresponds to an edge in the dependency graph) such that the matrix becomes lower triangular and can be solved in one pass through the matrix.
Such a pass through the cells of the system is called a transport sweep.

\subsection{Parallelization\label{sec:parallel}}
The sweep is clearly the most computationally intensive part of the source iteration algorithm (Algorithm~\ref{alg:sourceIteration}).
In order to apply this algorithm to large systems on modern hardware, we require some form of parallelization.
The easiest way of parallelizing this algorithm would be to distribute the solution of different frequencies $\nu$ and angles $\Omega$ onto the processors.
One problem with this method is that for very large numbers of processors there might simply not be enough different $\nu$ and $\Omega$ to efficiently employ all of them.
Moreover, parallelizing over $\Omega$ and $\nu$ requires the information about the grid to be present on every processor, which, due to memory requirements, quickly becomes infeasible for large simulations.
Due to these concerns, we choose to use a spatial decomposition of the grid.

For fully structured, euclidean grids, a sweeping algorithm and a domain decomposition that optimizes the parallel performance of the sweep is given by the Koch-Baker-Alcouffe algorithm~(\citet{bakerSnAlgorithmMassively1998}, \citet{kochSolutionFirstorderForm1991}).
This algorithm assumes that the number of processors $N$ can be factorized as $N = N_{\mathrm{x}} \cdot N_{\mathrm{y}}$. The domain is then subdivided $N_{\mathrm{x}}$ times along the $x$ axis and $N_{\mathrm{y}}$ times along the $y$ axis, resulting in a decomposition of the domain into $N$ columns. Each processor is then assigned one of the columns. For any direction $\Omega$, any column has at most three faces with upwind dependencies (it will have fewer dependencies only if $\Omega$ is aligned with one of the coordinate axes). If those upwind dependencies are fulfilled, the column can be solved in its entirety without further communication.
Of particular importance is that the solution of the upwind columns does not depend on incoming fluxes of the downwind column.
For unstructured grids, the latter statement does not hold, and a sweep can require many cycles of back-and-forth communication between neighbouring columns.
This makes the problem of finding the optimal domain decomposition for unstructured grids much harder.

In this work, we decide to use the already available domain decomposition in {\sc Arepo} (which is used for example for the hydrodynamics and gravity solvers) in order to simplify the problem and to reduce memory requirements.
The domain decomposition employed in {\sc Arepo} is based on the space-filling-curve approach.
The idea of this approach is to simplify the optimization problem by arranging all the cells of the three dimensional computational domain on a one dimensional line and then dividing that line into a number of domains with approximately equal estimated workload.
The advantage of using a space filling curve (such as the Peano-Hilbert curve) for this 1D to 3D mapping is that it results in reasonably localized domains (since the space filling curve maps points that are close in 1D to points that are close in 3D), thus reducing the amount of communication required.
In {\sc Arepo}, the estimated workload of a cell is given by a sum of the estimated work required for the gravitational and hydrodynamical calculations.
In principle, this estimate could be extended to include the workload due to radiative transfer, thus possibly reducing the overall time to solution by accelerating radiative transfer at the cost of a reduction in load balance for gravity and hydrodynamics.
However, for the sake of simplicity we choose not to do this in this work.

The remaining problem is to find an algorithm that performs the sweep across the entire grid which itself is distributed on different processors.
One challenge in this is that it is infeasible to calculate the topological ordering of the global dependency graph because this would require gathering the necessary information onto a single core or employing a parallel algorithm for topological sorting.
In the following section, we discuss our strategy for dealing with this problem, in which the topological ordering is never explicitly computed but instead implicitly adhered to by a task-based parallelism approach. This method is based on \citet{pautzAlgorithmParallelSweeps2002}.

\subsection{\label{sec:sweep}The sweep algorithm}
In the following, we define a task as a tuple $(c, \direc)$ of a cell $c$ and a sweeping direction $\direc$.
Solving a task means solving \eref{radiativeTransfer} in the cell $c$ for the direction $\direc$ for all frequencies $\nu$.
Note that we have excluded frequency from the definition of a task because we choose to solve all available frequencies at once whenever we solve a task.
For transport sweeps on structured grids, it is common to group the directions (for example into octants for a Cartesian grid) such that directions in the same group have the same dependency graph.
On an unstructured grid, two directions that are almost parallel can still have different dependency graphs, so we choose to do no grouping of the directions.

For any task $t = (c, \direc)$, we can define $d(t)$ / $u(t)$ to be the set of cells which are downwind / upwind of $c$ with respect to $\direc$. For a given grid cell at $\bm{r}$, both $d(t)$ and $u(t)$ can easily be obtained in a single loop through the neighbours, by counting a neighbour at position $\bm{r}_{\mathrm{n}}$ as downwind if $\left(\bm{r}_{\mathrm{n}} - \bm{r}\right) \cdot \bm{\Omega} > 0$ and as upwind otherwise. This operation can be done without any communication to other processors, since {\sc Arepo} ensures that grid cells belonging to other processors that are neighbours of any local cell are always present as local ghost particles and that the positions of the ghost particle is equivalent to the position of the corresponding cell on the other processor. Crucially, this ensures that the downwind/upwind neighbour relationship is always symmetric, even across processor boundaries.

With this, the unparallelized version of the algorithm to solve \stepref{sweep} in Algorithm~\ref{alg:sourceIteration} is given by~\ref{alg:serialSweep}.
Note that this algorithm requires non-periodic boundary conditions, which guarantees that at least one cell has $u(t) = 0$.
Extensions to periodic boundaries will be discussed in \secref{methodspbc}.
Since the dependency graph is acyclic, this algorithm will always terminate.

\begin{algorithm}
\caption{Single-core sweep}\label{alg:serialSweep}
\begin{algorithmic}[1]
\State initialize task queue $q \gets \bs{}$
\For{all $\direc$ and all cells $c$ in grid}
\State count number of required upwind fluxes $n(c, \direc) \gets |u(t)|$
\State\textbf{if} {$n(c, \direc) = 0$} \textbf{then} add task $(c, \direc)$ to $q$
\EndFor
\While{$q$ not empty}
\State get first task $t = (c, \direc)$ from $q$
\State solve $t$ using upwind fluxes \label{step:solver}
\For{downwind neighbor $c_{\mathrm{d}}$ in $d(t)$}
\State reduce missing upwind flux count $n(c_{\mathrm{d}}, \direc)$ by 1.
\State\textbf{if} {$n(c_{\mathrm{d}}, \direc)$ = 0} \textbf{then} add task $(c_{\mathrm{d}}, \direc)$ to $q$.
\EndFor
\EndWhile
\end{algorithmic}
\end{algorithm}

For cells on the boundary, which have no upwind dependencies, the incoming fluxes are obtained from the boundary conditions.
Fixed boundary conditions in which the value of the incoming radiation is $I_v = 0$ represent the simplest case.
In {\sc Arepo}, fixed boundaries are represented by cells with a connection to the first tetrahedron from which the grid was constructed and which encompasses the entire computational domain.
A discussion of periodic boundaries (represented by ghost cells which stand for a particle on the other side of the boundary) will follow in \secref{methodspbc}.

The exact way in which the radiative intensity is calculated from the upwind fluxes in \stepref{solver} will be discussed in \secref{solver}.

\begin{algorithm}
\caption{Parallel sweep}\label{alg:parallelSweep}
\begin{algorithmic}[1]
\State initialize task queue $q \gets \bs{}$
\State initialize send queues for each processor $i$ holding downwind neighbors of any of the cells in the domain of the current processor: $s_i \gets \bs{}$
\For{all $\direc$ and all cells $c$ in grid}
    \State count number of required upwind fluxes $n(c, \direc) \gets u(t)$
    \State \textbf{if} {$n(c, \direc) = 0$} \textbf{then} add task $(c, \direc)$ to $q$
\EndFor
\While{any cell unsolved or any $s_i$ not empty}
    \For{each incoming message (flux $f$ along $\direc$ into cell $c$)}
        \State reduce missing upwind flux count $n(c, \direc)$ by 1.
        \State \textbf{if} {$n(c, \direc)$ = 0} \textbf{then} add task $(c, \direc)$ to $q$.
    \EndFor
    \While{$q$ nonempty}
        \State get first task $t = (c, \direc)$ from $q$
        \State solve $t$ using upwind fluxes
        \For{downwind neighbor $c_{\mathrm{d}}$ in $d(t)$}
            \If{$c_{\mathrm{d}}$ is remote cell on processor $i$}
                \State add flux to send queue $s_i$ \label{step:sendQueue}
            \Else 
                \State reduce missing upwind flux count $n(c_{\mathrm{d}}, \direc)$ by 1.
                \State \textbf{if} {$n(c_{\mathrm{d}}, \direc)$ = 0} \textbf{then} add task $(c_{\mathrm{d}}, \direc)$ to $q$.
            \EndIf
        \EndFor
    \EndWhile
    \State send all messages in $s_i$
\EndWhile
\end{algorithmic}
\end{algorithm}

What we have described so far only works on a single processor.
In order to parallelize, we introduce Algorithm~\ref{alg:parallelSweep}, in which we communicate fluxes across processor domain boundaries.
Here, we had to make an implementation choice regarding the communication scheme.
The arguably simplest approach would be to send each flux immediately as we encounter it in \stepref{sendQueue}.
The benefit of this is that any downwind processor depending on the flux of this cell would be able to immediately obtain the required flux, thus potentially avoiding idle time.
In practice however, we found this approach to be too inefficient because of the communication delays it causes.
Therefore, we chose to buffer the fluxes in send queues and only send messages when there is nothing left to solve with the flux information we currently have.
This reduced the delays due to communication significantly and improved the scaling behavior in the idealized test cases.

Note that Algorithm~\ref{alg:parallelSweep} solves the sweep for different directions $\direc$ concurrently.
This is intentional, since it improves the parallel efficiency of the code.
If sweeps for different directions were performed in serial, processors with domains that are downwind in the direction of the sweep will be idle in the beginning of the sweep, while processors with domains that are upwind will be idle at the end. 

Note that a similar problem appears despite the parallel execution of different directions. It is called pipe fill or pipe drain~\citep{vermaakMassivelyParallelTransport2020}, and appears when the number of domains becomes large enough that there are inner regions which cannot start sweeping until outer regions are resolved. For an illustration of this effect, see \fref{pipeFill}, which shows a simplified case of a square-shaped domain decomposed into 16 subdomains.
As the figure shows, both the first and the last two directional sweeps will be performed while the central cores are idle, which reduces parallel efficiency.
As the number of cores grows, so does the duration of the pipe fill/drain phenomenon.
Note, that in Algorithm~\ref{alg:parallelSweep}, a partial sweep in a single direction is not necessarily finished before one in another direction is started, thus exacerbating the problem, compared to the scenario depicted in \fref{pipeFill}.
In addition to this, the domain decomposition and the dependency graph in~\fref{pipeFill} is much simpler than in an actual run of our code, due to the unstructured grid and the fact that the domain decomposition in our case has to be done with an eye towards the gravity and hydrodynamics solvers.

\begin{figure}
    \centering
    \def\svgwidth{\columnwidth}
    \fontsize{11pt}{31pt}\selectfont
\begingroup%
  \makeatletter%
  \providecommand\color[2][]{%
    \errmessage{(Inkscape) Color is used for the text in Inkscape, but the package 'color.sty' is not loaded}%
    \renewcommand\color[2][]{}%
  }%
  \providecommand\transparent[1]{%
    \errmessage{(Inkscape) Transparency is used (non-zero) for the text in Inkscape, but the package 'transparent.sty' is not loaded}%
    \renewcommand\transparent[1]{}%
  }%
  \providecommand\rotatebox[2]{#2}%
  \newcommand*\fsize{\dimexpr\f@size pt\relax}%
  \newcommand*\lineheight[1]{\fontsize{\fsize}{#1\fsize}\selectfont}%
  \ifx\svgwidth\undefined%
    \setlength{\unitlength}{1937.81925835bp}%
    \ifx\svgscale\undefined%
      \relax%
    \else%
      \setlength{\unitlength}{\unitlength * \real{\svgscale}}%
    \fi%
  \else%
    \setlength{\unitlength}{\svgwidth}%
  \fi%
  \global\let\svgwidth\undefined%
  \global\let\svgscale\undefined%
  \makeatother%
  \begin{picture}(1,0.75899321)%
    \lineheight{1}%
    \setlength\tabcolsep{0pt}%
    \put(0,0){\includegraphics[width=\unitlength,page=1]{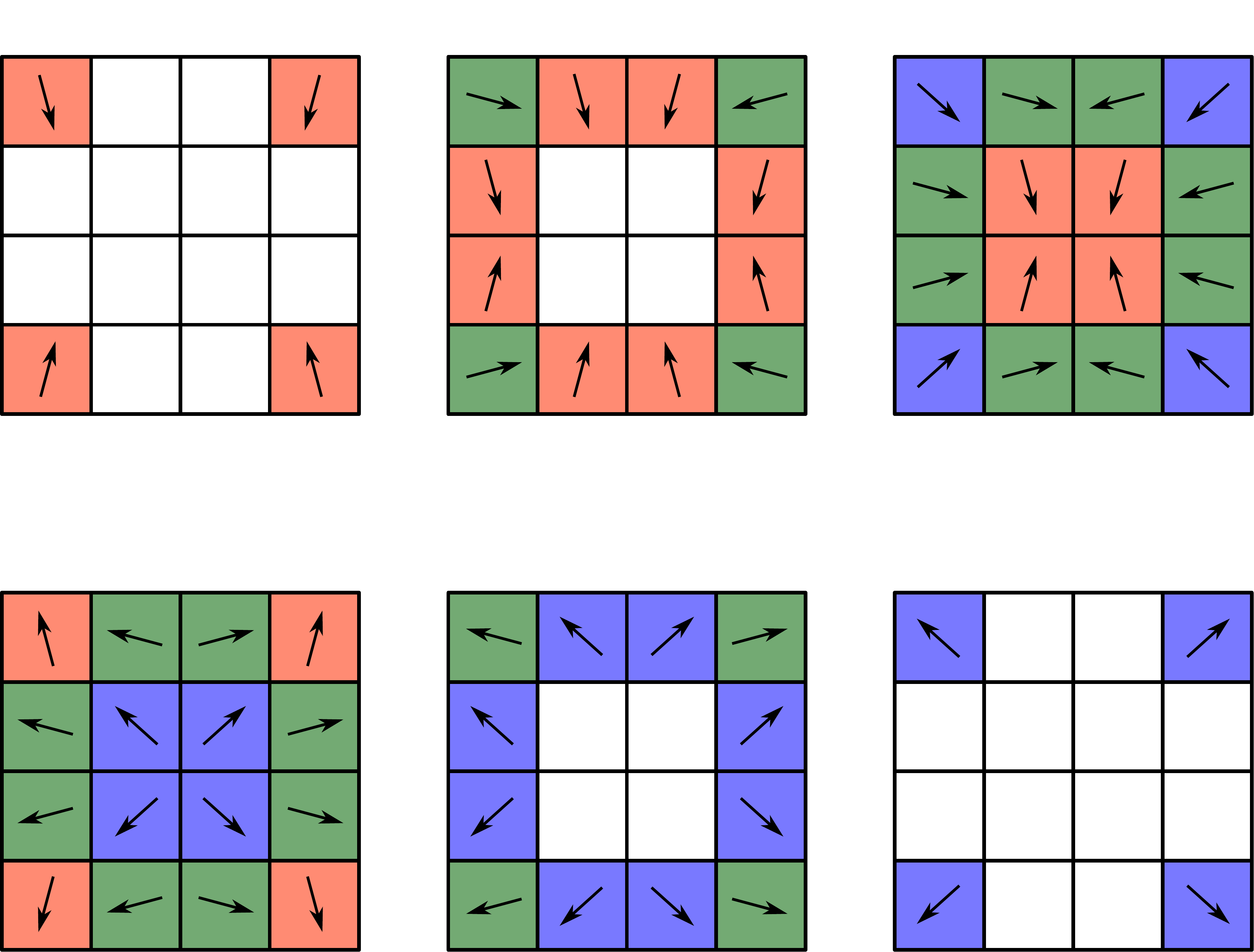}}%
    \put(0.42285814,0.74722744){\color[rgb]{0,0,0}\makebox(0,0)[lt]{\lineheight{1.25}\smash{\begin{tabular}[t]{l}Pipe Fill\end{tabular}}}}%
    \put(0.41273694,0.31491266){\color[rgb]{0,0,0}\makebox(0,0)[lt]{\lineheight{1.25}\smash{\begin{tabular}[t]{l}Pipe Drain\end{tabular}}}}%
  \end{picture}%
\endgroup%

    \caption{\label{fig:pipeFill} Illustration of the pipe fill/drain phenomenon. Each square denotes a computational domain belonging to a single processor. The arrows denote the direction of the sweep performed in that processor, while colors correspond to the (relative) time at which the sweep in that direction was first started, with red being before green which in turn denotes a time before blue. A white square without arrow means that the processor is currently idle.}
\end{figure}

One such problem which arises due to the unstructured grid is what we call re-entry dependencies. They appear when the sweep direction is close to being aligned with the boundary between two domains. In such cases, the dependencies can form a zig-zag pattern, such as the one depicted in \fref{reentry}. In such a scenario, the number of cells which can be solved locally before communication to the neighboring domain is required is very low. In the extreme, but not unrealistic, case depicted in \fref{reentry}, each processor can solve only one cell before having to communicate the resulting flux. While the effect is slightly alleviated by the fact that processors are not required to finish the solution of one sweep direction before starting the next, this still slows down the code significantly, mainly due to the additional delay each communication introduces. 

\begin{figure}
    \centering
    \def\svgwidth{\columnwidth}
    \fontsize{11pt}{31pt}\selectfont
\begingroup%
  \makeatletter%
  \providecommand\color[2][]{%
    \errmessage{(Inkscape) Color is used for the text in Inkscape, but the package 'color.sty' is not loaded}%
    \renewcommand\color[2][]{}%
  }%
  \providecommand\transparent[1]{%
    \errmessage{(Inkscape) Transparency is used (non-zero) for the text in Inkscape, but the package 'transparent.sty' is not loaded}%
    \renewcommand\transparent[1]{}%
  }%
  \providecommand\rotatebox[2]{#2}%
  \newcommand*\fsize{\dimexpr\f@size pt\relax}%
  \newcommand*\lineheight[1]{\fontsize{\fsize}{#1\fsize}\selectfont}%
  \ifx\svgwidth\undefined%
    \setlength{\unitlength}{153.9044051bp}%
    \ifx\svgscale\undefined%
      \relax%
    \else%
      \setlength{\unitlength}{\unitlength * \real{\svgscale}}%
    \fi%
  \else%
    \setlength{\unitlength}{\svgwidth}%
  \fi%
  \global\let\svgwidth\undefined%
  \global\let\svgscale\undefined%
  \makeatother%
  \begin{picture}(1,0.61039677)%
    \lineheight{1}%
    \setlength\tabcolsep{0pt}%
    \put(0,0){\includegraphics[width=\unitlength,page=1]{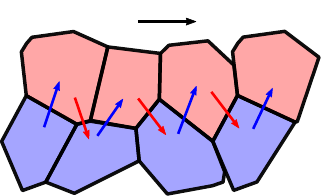}}%
    \put(0.49594314,0.56402516){\color[rgb]{0,0,0}\makebox(0,0)[lt]{\lineheight{1.25}\smash{\begin{tabular}[t]{l}$\bm{\Omega}$\end{tabular}}}}%
  \end{picture}%
\endgroup%

    \caption{\label{fig:reentry} Illustration of re-entry dependencies arising in scenarios where the sweep direction is aligned with the boundary between two domains. The red/blue color of the cells corresponds to the domain in which they reside. Blue arrows denote a dependency requiring inter-processor communication from the blue domain to the red, whereas red arrows denote communication from red to blue.}
\end{figure}

The problems described above can be solved partially by improved scheduling and communication strategies.
The main goal of such strategies is for the processors in the outer regions to solve the tasks required by those in the center as quickly as possible~\citep{adamsProvablyOptimalParallel2020} and to communicate the resulting fluxes immediately.
Such prioritization can greatly improve the parallel efficiency of this code by reducing the pipe-fill/drain effect.
For the sake of simplicity and to check whether this sweeping approach is feasible for the radiative transfer in {\sc Arepo}, in this paper, we use a very simple prioritization strategy which prioritizes finishing one sweep direction before starting another one.

Another possible optimization is to intentionally omit certain fluxes between cells, thus removing dependencies from the graph. Doing so means that the result of the transport sweep is only an approximation and obtaining the solution would require iterating over a number of sweeps. However, if the right dependencies are removed (e.g. the re-entry dependencies discussed above), the performance improvement can potentially be large enough to offset the additional cost due to the iteration~\citep{lucerolorcaMultilevelSchwarzMethods2018}.

\subsection{Transport methods\label{sec:solver}}
In order to calculate the (downwind) fluxes out of a cell, given the source terms, absorption coefficients and the incoming (upwind) fluxes, we need to decide on a transport scheme with which we can solve \eref{radiativeTransfer}.

\begin{figure}
    \centering
    \def\svgwidth{\columnwidth}
    \fontsize{13pt}{31pt}\selectfont
\begingroup%
  \makeatletter%
  \providecommand\color[2][]{%
    \errmessage{(Inkscape) Color is used for the text in Inkscape, but the package 'color.sty' is not loaded}%
    \renewcommand\color[2][]{}%
  }%
  \providecommand\transparent[1]{%
    \errmessage{(Inkscape) Transparency is used (non-zero) for the text in Inkscape, but the package 'transparent.sty' is not loaded}%
    \renewcommand\transparent[1]{}%
  }%
  \providecommand\rotatebox[2]{#2}%
  \newcommand*\fsize{\dimexpr\f@size pt\relax}%
  \newcommand*\lineheight[1]{\fontsize{\fsize}{#1\fsize}\selectfont}%
  \ifx\svgwidth\undefined%
    \setlength{\unitlength}{312.44850603bp}%
    \ifx\svgscale\undefined%
      \relax%
    \else%
      \setlength{\unitlength}{\unitlength * \real{\svgscale}}%
    \fi%
  \else%
    \setlength{\unitlength}{\svgwidth}%
  \fi%
  \global\let\svgwidth\undefined%
  \global\let\svgscale\undefined%
  \makeatother%
  \begin{picture}(1,0.42515982)%
    \lineheight{1}%
    \setlength\tabcolsep{0pt}%
    \put(0,0){\includegraphics[width=\unitlength,page=1]{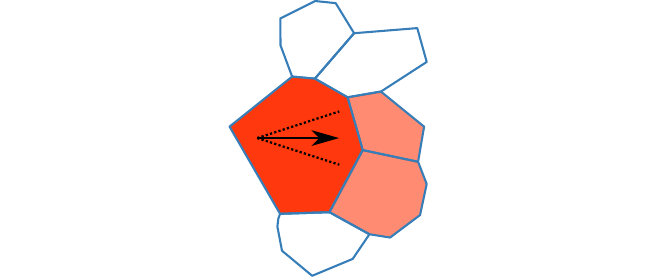}}%
    \put(0.356615,0.34365526){\color[rgb]{0,0,0}\makebox(0,0)[lt]{\lineheight{1.25}\smash{\begin{tabular}[t]{l}b)\end{tabular}}}}%
    \put(0,0){\includegraphics[width=\unitlength,page=2]{transportSchemes.pdf}}%
    \put(0.69679567,0.34365526){\color[rgb]{0,0,0}\makebox(0,0)[lt]{\lineheight{1.25}\smash{\begin{tabular}[t]{l}c)\end{tabular}}}}%
    \put(0,0){\includegraphics[width=\unitlength,page=3]{transportSchemes.pdf}}%
    \put(0.01147336,0.34365526){\color[rgb]{0,0,0}\makebox(0,0)[lt]{\lineheight{1.25}\smash{\begin{tabular}[t]{l}a)\end{tabular}}}}%
    \put(0,0){\includegraphics[width=\unitlength,page=4]{transportSchemes.pdf}}%
  \end{picture}%
\endgroup%

    \caption{\label{fig:transportSchemes}Illustration of the three different transport schemes.
      The black arrow represents the sweeping direction $\bm{\Omega}$.
      The dotted lines represent the solid angle corresponding to the direction $\bm{\Omega}$.
      The shading of the downwind cells represents the flux that the cells would receive, with white meaning no incoming flux and red meaning a high amount of incoming flux.
      a) Distribution proportional to fraction of area of the cell interfaces to the total area; b) Choosing the $n$ most straightforward neighbors; c) Choosing the $n$ most straightforward neighbors along a random vector in the solid angle corresponding to the direction.}
\end{figure}

Three such schemes are depicted in \fref{transportSchemes}.
Note that the solid angle corresponding to a given direction is given by $\frac{4 \pi}{N_{\mathrm{dir}}}$ where $N_{\mathrm{dir}}$ is the number of directions in our discretization.

In the first scheme, in Panel a), the outgoing flux $F_i$ of radiation for the direction $\direc$ to a downwind neighbor $i$ is given according to the distribution
\begin{align}
  F_i = F \frac{A_i \bb{\bm{n}_i \cdot \direc}}{\sum_{j=1}^N A_j \bb{\bm{n}_j \cdot \direc}},
\end{align}
where $F$ is the total outgoing flux (given by the sum of non-absorbed incoming radiation and the radiation created by the source term of this cell), $\bm{n}_i$ and $\bm{n}_j$ are the normals of the Voronoi faces connecting the cell to neighbours $i$ and $j$ respectively and $A_i$ and $A_j$ are the areas of the faces.

The Simplex2 method~\citep{kruipMathematicalPropertiesSimpleX2010,paardekooperSimpleX2RadiativeTransfer2010}, which is the basis for the SPRAI implementation in {\sc Arepo}~\citep{jauraSPRAICouplingRadiative2018}, introduces an additional transport method (called direction-conserving transport) in which the incoming flux is distributed equally onto $n$ neighbors with the most straight-forward face normals along $\direc$, see Panel b) in \fref{transportSchemes}. The authors showed that $n = \ndim$, with $\ndim$ being the number of spatial dimensions is the optimal choice for direction-conserving transport.
The idea of this scheme is to reduce numerical diffusion.
However, this comes at the cost of amplifying the effect that the angular discretization into a number of discrete directions introduces, namely that radiation is transported along preferential directions~\citep{jauraSPRAICouplingRadiative2018}, something that becomes very apparent in optically thin media where the mean free path is long.
In principle, this behavior could be alleviated by increasing the number of directions.
However, this increases memory requirements and computation time.
In SPRAI this problem is solved in two ways.

Firstly, a slightly modified version of the direction conserving transport is employed in which the direction in which radiation will be transported is decided on a cell-by-cell basis. For each cell, instead of transporting radiation along $\direc_i$, a vector $\direc_i^{\prime}$ is chosen randomly, with the only condition being that $\direc_i^{\prime}$ is closer to $\direc_i$ than any of the angles $\direc_j$ for $i \neq j$ (in other words, $\direc_i^{\prime}$ should be within the solid angle that $\direc_i$ corresponds to). This method is illustrated in Panel c) in \fref{transportSchemes}.
We choose not to implement this transport method for sweep, because it would require us to implement the random choice of $\direc^{\prime}$ in a deterministic fashion, in order to allow us to properly count the number of upwind/downwind dependencies.
The drawback of this is that our results will not agree exactly with those of SPRAI, even in the absence of any scattering, because of the different choice of transport method.

The second way in which SPRAI reduces preferential directions is that any radiative transfer step may be subdivided into $\Nrot$ substeps, each with the source terms reduced by a factor of $1 / \Nrot$.
For each step, the radiation chemistry is updated according to the resulting intensity field.
After every substep, the directions $\bm{\Omega}_i$ are rotated to new directions $\bm{\Omega}_i = \bm{R}(\theta, \phi) \cdot \bm{\Omega}_i'$ where $\bm{R}(\theta, \phi)$ is the rotation matrix and the spherical coordinate-angles  $\theta$ and $\phi$ are randomly chosen as $\theta \in [0, \pi]$, $\phi \in [0, 2 \pi]$.
The remapping between angle-dependent quantities, such as the intensity is then done via $I_{\nu}(\bm{r}, \bm{\Omega}_i') = \sum_{j=1}^{N_{\mathrm{dir}}} c_{i j} I_{\nu}(\bm{r}, \bm{\Omega}')$ where $N_{\mathrm{dir}}$ is the number of discrete directions and the interpolation coefficients $c_{i j}$ depend on the choice of interpolation and obey $\forall i: \sum_{j=1}^{N_{\mathrm{dir}}} c_{i j} = 1$
For simplicity, we choose $c_{i j} = \frac{\Delta S_{i j}}{\Delta S_i}$, where $\Delta S_{i j}$ is the solid angle that $\bm{\Omega}_i$ and $\bm{\Omega}_j$ share and $\Delta S_i$ is the solid angle corresponding to any direction $\Omega_i$.
This random rotation of the directions effectively smears out preferential directions at the cost of additional computation time.
In SPRAI, radiation travels one cell at a time before the scattering terms are re-computed.
This process is repeated until all photons are absorbed.
Throughout this, SPRAI needs the directions to remain constant (otherwise, direction would not be conserved for more than a cell length).
In the sweep method, the directions only need to remain constant throughout one single sweep. This means we can combine the source iteration (Algorithm~\ref{alg:sourceIteration}) and the rotation of the directions, potentially saving many iterations.

In all our tests, we use the first transport method in which outgoing fluxes are simply assigned via the geometry of the cell.
A potential benefit of the direction conserving transport method is that it reduces the average number of downwind dependencies per cell from ${15.54}/{2} \approx 7.8$ (since the average number of neighbors in a 3D-Voronoi grid is $15.54$) to the number of dimensions, $\ndim = 3$, thus making the dependency graph thinner.

\subsection{Periodic Boundary Conditions\label{sec:methodspbc}}
In simulations of the period of reionization, the simulated volume is often selected as a box which is supposed to be representative of the universe. During a normal simulation of such a box using gravity and hydrodynamics, periodic boundary conditions are employed to effectively model the influence of the adjacent regions of space without having to simulate those regions explicitly. The same idea applies to the radiative transfer. Using periodic boundaries, any photons leaving the box can re-enter it from the opposing side. If the box is large enough to be statistically representative, then this re-entry models the light from the neighboring regions.

In order to introduce periodic boundary conditions in simulations, the standard approach is to add a mirror image of each boundary cell on the other side of the grid, i.e. to add the same cell with its position shifted by the box size $L$. These mirror images are called ghost cells in Arepo. Fluxes going into such a ghost cell will then be treated as incoming fluxes into the corresponding \emph{normal} cell.

\begin{figure}
    \centering
    \def\svgwidth{\columnwidth}
    \fontsize{13pt}{31pt}\selectfont
    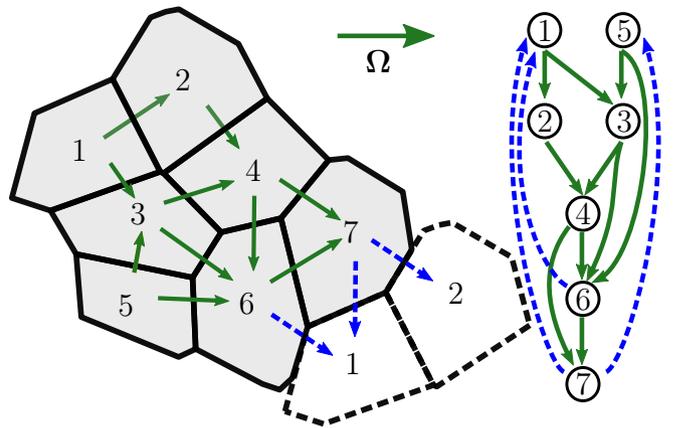
    \caption{\label{fig:gridCellsPeriodic}Left: Illustration of a 2D Voronoi grid and the dependencies induced by the sweep ordering for a sweep towards the right under the assumption of periodic boundary conditions. Solid boundaries and gray background represent normal cells, dashed boundaries and white background represent periodic ghost cells. Right: The directed, cyclic graph corresponding to the dependencies.}
\end{figure}

In transport sweeps, introducing such ghost cells at the boundaries poses an additional challenge. As illustrated in \fref{gridCellsPeriodic}, after the inclusion of the ghost cells, the induced dependency graph becomes cyclic. Clearly, no topological ordering exists for cyclic graphs. Simply applying Algorithm~\ref{alg:parallelSweep} to a grid with periodic boundaries thus cannot work - the algorithm would never terminate.

In order to solve this, we use an iterative approach, similar to Alg.~\ref{alg:sourceIteration}. Any radiation going out of the boundaries of the computational domain is added to the effective source term of its periodic ghost cell for the next iteration. This breaks the cyclic dependencies induced by the periodic boundary condition but still ensures that any outgoing radiation is re-introduced into the box.

There are numerous choices for how to define convergence for this iterative method. We use the relative difference in the source terms $j_i^n$ (where $i$ denotes the cell and $n$ denotes the iteration number) as an error
\begin{equation}
  \label{eq:relativeErrorPbc}
E_i^n = \frac{j_i^n - j_i^{n-1}}{j_i^n + j_i^{n-1}}.
\end{equation}
The iteration is stopped if $\forall i: E_i < \epsilon$, where $\epsilon$ is the convergence threshold which can be chosen by the user. Additionally, we define a maximum number of iterations after which the algorithm will terminate, even if there are still cells which exceed the error threshold. 

Clearly, if the number of iterations needed in order to reach convergence in the source terms of the periodic boundary iteration is $\nItPbc$, while the number of iterations needed to relax the terms introduced due to scattering (Algorithm~\ref{alg:sourceIteration}) is $\nItScat$, this iterative scheme increases the overall runtime of the algorithm by a factor of $\nItPbc \nItScat$ compared to the runtime of a single sweep.
In order to improve on this, an interesting approach could be to combine the two iterative schemes, such that the source terms due to scattering and the periodic boundary conditions are calculated at the same time.
If doing so does not change the behavior of the individual schemes, this would reduce the runtime overhead to $\mathrm{max}(\nItPbc, \nItScat)$.

As a way to reduce $\nItPbc$, we tried an approach we call ``warm starting'' in which the final values of the source terms obtained in a previous RT step are used as an initial guess in the next step, instead of using $j_i^n = 0$ as a guess. 
This is made technically challenging by the fact that the grid might change between one RT step and the next, for example by removing cells from the computation or by introducing new ones in adaptive refinement schemes.  As a guess for the source term for any newly created cell we use $j_i^n = 0$.

\section{Tests}\label{sec:tests}
For all the test simulations we make the following choices regarding the parameters of the numerical discretization.
For the frequency discretization, we choose a single frequency bin corresponding to photons in the range $\bb{\SI{13.6}{\electronvolt}, \infty}$, i.e.\ with enough energy to ionize hydrogen.\footnote{Note that Sweep can readily deal with multiple energy bins; we make this choice here purely for simplicity.}
For the angular discretization, we use \num{84} directions isotropically distributed over the unit sphere generated by simulated annealing \citep{jauraSPRAICouplingRadiative2018}.
The code supports other numbers of directions but we choose \num{84} as a compromise between lower numbers which reduce the accuracy of the solution and higher numbers which increase memory consumption and overall runtime and to be consistent with the results of \citet{jauraSPRAICouplingRadiative2018}, where the number of directions was also chosen to be \num{84}.

We use $\Nrot = \num{5}$ random rotations of the directions in every time step.
We find that this number of rotations is sufficient to smooth out any obvious preferential directions in the results and still small enough to keep the run time reasonable.

For all tests, we fix the hydrogen ionization cross section at $\sigma_{\mathrm{H}} = \SI{5.38e-18}{\cm\squared}$, corresponding to the value for a 14.4~eV photon, and the case B recombination rate coefficient to a constant value $\alphaB = \SI{2.59e-13}{\centi\metre\cubed\per\second}$.

\subsection{Expansion tests\label{sec:expansionTests}}
We consider first the idealized scenario of an ionizing source surrounded by a homogeneous distribution of neutral atomic hydrogen. Here, the source will form a spherical region of ionized hydrogen around it, known as an HII region.
\citet{stromgrenPhysicalStateInterstellar1939} showed that in ionization equilibrium, the radius of this region is given by the Strömgren radius,
\begin{align}
  \Rstroemgren = \bb{\frac{3 N_{\gamma}}{4 \pi \alphaB \nelectron n_{\rm H^{+}}}}^{1/3},
\end{align}
where $\alphaB(T)$ is the case B recombination coefficient of hydrogen and $\nelectron$ is the electron number density. Given that the medium inside the spherical region is highly ionized, it follows that $\nelectron \approx n_{\rm H^{+}}$.

In the initial phase of the evolution, the expansion is simply driven by radiation which ionizes the neutral gas just beyond the ionization front (I-front).
It takes place at very high velocities, compared to the speed of sound in the ionized gas $\cs$, so that the hydrodynamical response of the gas is irrelevant for the movement of the ionization front.
This initial, rapid expansion is called the R-type expansion (R=rarefied).

Under the assumption that the density of the gas remains constant, the time-evolution of the radius of the ionization sphere is given by
\begin{align}
  \Rr(t) = \Rstroemgren \bb{1 - e^{-t / \trec}}^{1/3},
  \label{eq:rtypeRadius}
\end{align}
where $\trec = (\alphaB \nH)^{-1}$ is the recombination time.

Once the radius of the sphere reaches the Strömgren radius, the second phase of the evolution, called D-type (D=dense) begins.
In this phase, the expansion of the sphere is driven by a pressure gradient between the ionized, inner region and the neutral, outer region.
This pressure gradient is caused by the large temperature difference between the two regions.
In this second phase, the I-front is preceded by a shock front since it moves at velocities that are supersonic in the neutral medium but subsonic in the ionized medium.
An analytical expression for the radius of the sphere as a function of time was first derived by \citet{spitzerPhysicalProcessesInterstellar1978} and is given by
\begin{align}
  \Rd(t) = \Rstroemgren \bb{1 + \frac{7}{4} \frac{\cs t}{\Rstroemgren}}^{4/7},
  \label{eq:dtypeRadius}
\end{align}
where $t = 0$ here corresponds to the time at which the ionization front transitions from R-type to D-type.

\subsubsection{R-type expansion\label{sec:rtype}}
As a first test of the radiative transfer code, we study the R-type expansion of a HII region. 
The following tests are performed at 3 different resolutions of $32^3$, $64^3$, and $128^3$ cells.
We use the same initial conditions as those in the R-type expansion test in~\cite{jauraSPRAICouplingRadiative2018} and~\cite{baczynskiFerventChemistrycoupledIonizing2015}, in order to compare our results.
The simulation box is a cube with side length $L = \SI{12.8}{\kilo\parsec}$.
At the center of the box is a idealized point source which emits photons at a rate of $\dot{N}_{\gamma} = \SI{1e49}{\per\second}$.
The box is initialized with a gas with homogeneous number density $\nH = \SI{e-3}{\per\centi\metre\cubed}$.
With these parameters, we find values of $\Rstroemgren = \SI{6.79}{\kilo\parsec}$ for the Strömgren radius and $\trec = \SI{122.4}{\mega\year}$ for the recombination time. 
We initialize the gas as being purely neutral hydrogen (i.e. $x_{\mathrm{H}} = 1$, $x_{\mathrm{H+}} = 0$). 
Since the density response of the gas is irrelevant for the R-type expansion, we run the simulation without hydrodynamics, so that only radiative transfer and ionization chemistry take place.

In order to compare the time evolution of the radius of the ionized sphere to the analytical prediction, we need to define the radius of the sphere.
In the simple analytical model, there is a sharp transition between the ionized and the non-ionized regions.
However, in our simulation, due to the limited resolution of the grid, the transition region has a finite size. 
This means that a different definition of the radius of the sphere is required.
Here, we define the radius $R(t)$ as the radius at which the average ionization is $x_{\rm H+} = x_{\rm H} = 0.5$, i.e.
\begin{align}
  \label{eq:definitionExpansionRadius}
  \int_{\bm{S}(R)} \mathrm{d}\bm{r} x_{\rm H}(\bm{r}) = 0.5,
\end{align}
where $\bm{S}(R)$ denotes the surface of the sphere of radius R around the origin.
To calculate the value of this integral in practice, we average the HII abundance over a spherical shell of a given thickness $\Delta \ll R$.

\begin{figure}
    \includegraphics{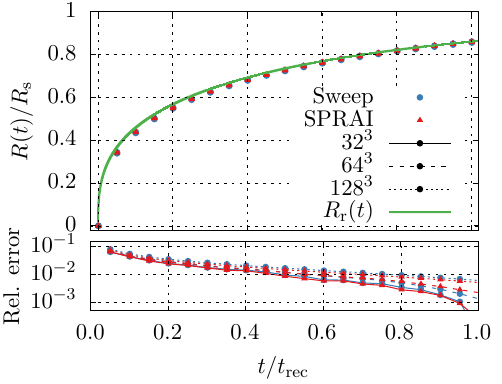}
    \caption{\label{fig:rtype}R-type expansion of a ionization front in a uniform medium. Top panel: Radius $R(t)$ of the ionization sphere normalized by the Strömgren radius $\Rstroemgren$ as a function of time $t$, normalized by the recombination time $\trec$. Blue dots: Numerical results for Sweep. Red triangles: Numerical results for SPRAI, Solid lines: Results for $32^3$ particles, Dashed lines: Results for $64^3$ particles, Dotted lines: Results for $128^3$ particles. Green line: Analytical prediction $\Rr(t)$ given by \eref{rtypeRadius}. Bottom panel: Relative error $\abs{R(t) - \Rr(t)} / \Rr(t)$}
\end{figure}

In the upper panel of \fref{rtype}, the radius $R(t)$ (normalized by the Str\"omgren radius $\Rstroemgren$) of the ionized sphere is shown as a function of time (normalized by the recombination time $\trec$).
In the lower panel, the relative error of the results compared to the analytical prediction is shown as a function of time.
The results are shown for the three resolutions. For each resolution, we also show a comparison to the results obtained by performing the same simulation with the SPRAI code, as well as to the analytical prediction given by \eref{rtypeRadius}.

The comparison of analytical prediction and the simulation results shows that, while after the first timestep, the error is on the order of $\approx 8\%$, it decreases with time and drops below $1\%$ for all resolutions towards the end of the simulation.
In contrast to our expectations, the agreement with the analytical prediction decreases with increasing resolution.
While the simulation with $128^3$ particles shows a relative error of $\approx 0.8 \%$ at the end of the simulation, the simulation with $32^3$ particles drops to an error of $\approx 0.1 \%$ at the same time.
We do not have an intuitive explanation for these results.
However, we emphasize that the analytical prediction assumes a perfectly sharp boundary, which does not exist in practice, where the boundary has an associated thickness.
Due to this, the value of the radius depends quite strongly on the definition of the radius in \eref{definitionExpansionRadius}. While the choice of a ionization threshold $0.5$ is intuitive, a different value will give rise to different radii and therefore change the dependence of the relative error on the resolution of the simulation.

For all resolutions, the results of Sweep and SPRAI agree very well, which increases our confidence in the numerical results.
There is no clear difference between the relative errors of the two codes.
While the relative error is slightly lower for Sweep at $128^3$, the exact opposite is visible at $64^3$ where SPRAI shows slightly lower errors.
At $32^3$, the results of both codes agree well with the analytical prediction and show virtually no difference in the relative error.

\subsubsection{D-type expansion}
Our D-type expansion test is set up very similar to the R-type test.
The main qualitative difference between the two setups is that we need to take hydrodynamics into account, since the D-type expansion is due to the gas response driven by the thermal pressure between the inner, ionized region and the outer, neutral region.
As in the R-type test, we chose our parameters as in \cite{jauraSPRAICouplingRadiative2018}, in order to facilitate comparison.
We perform the D-type expansion for the $128^3$ resolution case.

\begin{figure}
    \includegraphics{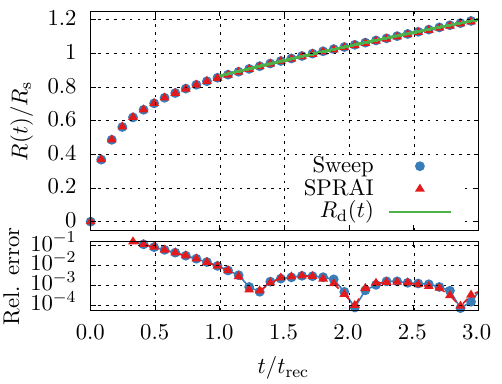}
    \caption{\label{fig:dtype}D-type expansion of a ionization front in a uniform medium. Top panel: Radius $R(t)$ of the ionization sphere normalized by the Strömgren radius $\Rstroemgren$ as a function of time $t$, normalized by the recombination time $\trec$. Blue dots: Numerical results for Sweep. Red triangles: Numerical results for SPRAI, Green line: Analytical prediction $\Rd(t)$ given by \eref{dtypeRadius}. Bottom panel: Relative error $\abs{R(t) - \Rd(t)} / \Rd(t)$}
\end{figure}

In order to use the analytical prediction given by \eref{dtypeRadius}, we need to obtain a value for the speed of sound in the ionized medium $\cs$.
In principle, one could obtain the speed of sound using the temperature of the ionized medium via
\begin{equation}
  \cs = \sqrt{\frac{\gamma k_{\mathrm{B}} T_{\mathrm{avg}}}{\mu m_{\mathrm{H}}}},
\end{equation}
where $\gamma = \frac{5}{3}$ is the adiabatic index, $\mu$ is the mean molecular weight (in atomic units) and $m_{\mathrm{H}}$ is the atomic mass of hydrogen.

However, this is difficult in practice, since the temperature is not constant inside the ionized sphere.
Therefore, we treat the speed of sound $\cs$ as a fit parameter to our data, which is consistent with the approach in~\cite{jauraSPRAICouplingRadiative2018}.
In doing so, we obtain a value of $\SI{12.8}{\km\per\s}$ corresponding to an average temperature $T_{\mathrm{avg}} = \SI{11914}{\K}$.

In the top panel \fref{dtype}, the dependence of the radius of the ionization sphere, normalized by the Strömgren radius $\Rstroemgren$ is shown as a function of the time, normalized by the recombination time $\trec$ as well as the analytical prediction given by \eref{dtypeRadius}. The prediction describes the behavior for $R(t) > \Rstroemgren$, but we also display the solution at lower times, starting at $t = \trec$ and find that it describes the data quite well even in this range.
This is confirmed by the relative error of the data with respect to the analytical prediction, shown in the lower panel of \fref{dtype}. Beginning at $t = \trec$ the error never exceeds $1 \%$. We find no discernible difference in the relative error between the results of Sweep and SPRAI, solidifying the fact that Sweep produces physically correct results.

Since this is the only test involving hydrodynamics, we will also discuss the relative performance of radiative transfer compared to the other parts of the code here, even though it should be noted that such a performance comparison is problem dependent.
In a run using Sweep, the radiative transfer takes up approximately $75 \%$ of the total computation time, with Voronoi grid construction ($12 \%$) and hydrodynamics ($10 \%$) using up most of the remaining time.
While this implies that in this test radiative transfer is by far the most expensive part of the code, Sweep still vastly outperforms SPRAI (which takes up $\approx 98 \%$ of the total run time) by a factor of $\approx 16$.

\subsection{\label{sec:shadowing}Shadowing behavior of radiation field behind a clump}
\begin{figure*}
    \includegraphics{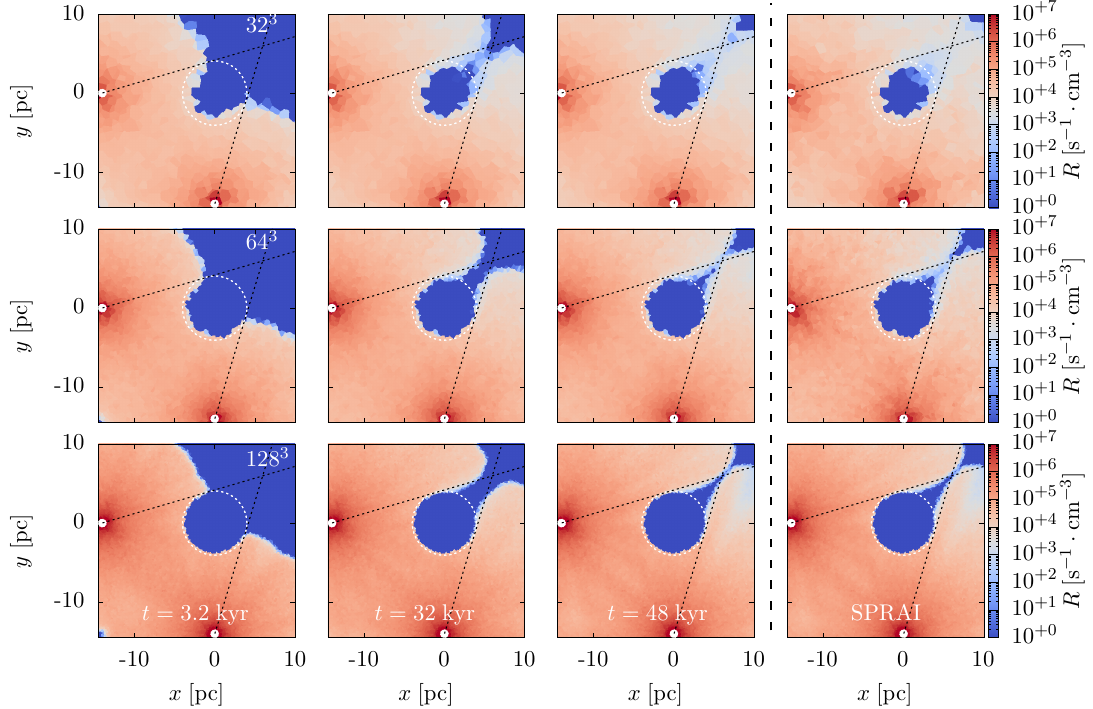}
    \caption{\label{fig:shadowing}The photon rate $R$ in a slice through the $z=0$-plane of the simulation box.
      First row: $32^3$ particles,
      Second row: $64^3$ particles,
      Third row: $128^3$.
      First column: Sweep at $t = \SI{3.2}{\kilo\year}$,
      second column: Sweep at $t = \SI{32}{\kilo\year}$,
      third column: $t = \SI{48}{\kilo\year}$,
      last column: SPRAI at $t = \SI{48}{\kilo\year}$.
      The white dashed line represents the over-dense clump.
      White solid circles represent the position of the sources.
      The black dashed lines delineate the shape of an ideal shadow behind the clump.
    }
\end{figure*}

The previous tests have established that the Sweep method replicates results obtained with SPRAI.
However, due to the spherical symmetry, the directional dependence of the radiation is not tested in the expansion tests.
To do so, we perform another test in which we study how well dense objects cast shadows behind them.

This test is set up in the same way as the corresponding test in~\cite{jauraSPRAICouplingRadiative2018} and we will use the results obtained by the SPRAI method as a basis for comparison.
The simulation takes place in a box of side length $L = \SI{32}{\parsec}$, filled with neutral hydrogen at a number density of $\nH = \SI{1}{\per\cubic\centi\metre}$ everywhere except in the center of the box where a dense clump at number density 
$\nH = \SI{1000}{\per\cubic\centi\metre}$ and radius $R = \SI{4}{\parsec}$ is placed.
The temperature of the gas is set to $T = \SI{1000}{\kelvin}$.
Two point sources are placed at $\bm{r}_1 = (-14, 0, 0) \; \mathrm{pc}$ and $\bm{r}_2 = (0, -14, 0) \; \mathrm{pc}$, both emitting photons at a rate of $N_{\gamma} = \SI{1.61e48}{\per\second}$. 
The time-step of the simulation is $\Delta t = \SI{0.32}{\kilo\year}$.

An analysis of this test, which includes hydrodynamics and discusses the temperature, pressure, and density response has been performed in the original SPRAI paper \citep{jauraSPRAICouplingRadiative2018}.
Since the code coupling the radiative transfer to the hydrodynamics of Arepo is the same as the one used in SPRAI, any results obtained there are also valid for our method.
Since we are interested only in the photon rate field resulting from the simulations, we perform these simulations without hydrodynamics.
Here, the photon rate $R(\bm{r}, t)$ is defined as the number of photons in the frequency bin corresponding to the ionization of hydrogen at $\SI{13.6}{eV}$ per unit time per unit volume, i.e. as $R(\bm{r}, t) = \frac{1}{\SI{13.6}{eV}} \int_{\bm{\Omega}} I_{\nu}(\bm{r}, \bm{\Omega}, t) \mathrm{d}\bm{\Omega}$.
In \figref{shadowing}, the photon rate $R$ is shown as a slice through the simulation box along the x-y plane for different times (columns) and resolutions (rows). For each resolution, the result obtained with SPRAI is shown for the last time (i.e. $t = \SI{48}{\kilo \year}$).

It is clear that the over-dense clump acts as an obstacle and initially prevents photons from entering its shadow.
However, due to numerical diffusion, the shadow is not as sharp as expected in the exact solution.
As time progresses, the photon rate in the (theoretical) shadow behind the clump increases, because the regions between the sources and the shadow have become ionized and stopped absorbing photons.
With increasing resolution, the effect of numerical diffusion decreases and the shadow becomes more defined.

In order to quantify the shadowing behavior and to compare Sweep and SPRAI as well as the quality of the shadow at different resolutions, we calculate the mass averaged fraction of ionized hydrogen in the volume of the shadow.
The volume is given by the intersection of two (infinitely extended) cones, with their tips at $\bm{r}_1$ and $\bm{r}_2$ respectively and their base determined by the great circle lying in the over-dense clump. In the 2D slice shown in \fref{shadowing}, this volume $V_{\rm S}$ corresponds to the area between the white dashed circle and the black dashed lines.
This fraction $\bar{x_{\rm H}}$ is given by
\begin{align}
  \label{eq:volumeIntegral}
  \overline{x_{\rm H}} = \frac{\int_{V_{\rm S}} x_{\rm H}(\bm{r}) \rho(\bm{r})  \mathrm{d}V}{\int_{V_{\rm S}}  \rho(\bm{r})  \mathrm{d}V},
\end{align}
where $x_{\rm H}(\bm{r})$ is the abundance of ionized hydrogen at position $\bm{r}$ and $\rho(\bm{r})$ denotes the mass density at position $\bm{r}$.

\begin{figure}
    \includegraphics{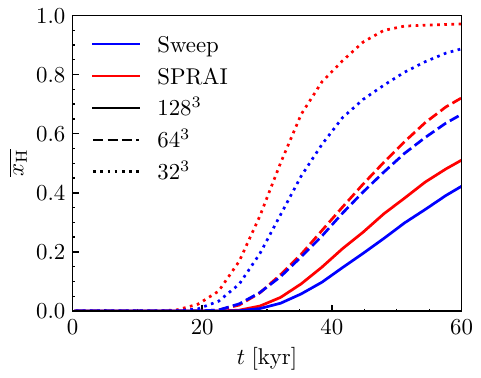}
    \caption{\label{fig:shadowingVolume}
      The average hydrogen abundance $\overline{x_{\rm H}}$ (see \eref{volumeIntegral}) in the shadow volume as a function of time for both Sweep (blue) and SPRAI (red) for three different resolutions:
      $128^3$ (solid line),
      $64^3$ (dashed line) and
      $32^3$ (dotted line),
    }
\end{figure}

In \figref{shadowingVolume}, $\overline{x_{\rm H}}$ is shown as a function of time. 
Neither Sweep nor SPRAI form a perfect shadow, demonstrated by the fact that the ionization fraction begins to increase at $t \approx \SI{20}{\kilo \year}$. Before this time, the ionization front has not reached the region behind the over-dense clump. 
Clearly, the shadowing behavior improves with higher resolutions. This is in line with the explanation that the protrusion of the ionization front into the shadow is due to numerical diffusion, since higher resolutions decrease the effect of numerical diffusion.
We also find that Sweep forms a slightly more defined shadow, with the ionization fraction strictly below the values for SPRAI for all times and resolutions.

\subsection{Scattering\label{sec:scattering}}
In order to test the source iteration scheme described in Section~\sref{sourceIteration}, we test a setup similar to the illumination of a dense clump described in Section~\sref{shadowing}. The only difference is that this test setup will only use one source positioned at $\bm{r}_1 = (-4.8, 0, 0) \; \mathrm{pc}$, which is very close to the dense clump positioned in the center (which has radius $r = \SI{4}{pc}$), creating a large shadow behind the clump.

In order to test that our implementation of the source iteration reproduces scattering in a physical manner, we perform a number of simulations in which we vary only the effective scattering cross section.
For simplicity, we choose a model in which the scattering coefficient is entirely independent of the chemical composition of the gas, with the scattered fraction of the intensity in a cell given by
\begin{equation}
  \bb{\mathrm{d}I_{\nu}}_{\mathrm{s}} = I_{\mathrm{\nu}} \bb{1-e^{-\mathrm{d}n_{\mathrm{nucleons}} \sigmaS}},
\end{equation}
where $_{\mathrm{\nu}}$ is the incoming intensity, $n_{\mathrm{nucleons}}$ is the column density of nucleons in the cell, $\sigmaS$ is the effective cross section of the scattering.
The column density is calculated as
\begin{equation}
  \mathrm{d}n_{\mathrm{nucleons}} = n \langle \mathrm{d} r \rangle,
\end{equation}
with the number density of nucleons $n$ and the mean distance traveled in the cell $\langle \mathrm{d} r \rangle$.
For more details of these calculations see Section 2.2 in~\cite{jauraSPRAICouplingRadiative2018}

For the resolution of the tests, we chose $n = 128^3$ particles for all test simulations.
We vary the scattering cross section as $\sigmaS =
\SI{0}{\per\cm\squared},
\SI{5e-22}{\per\cm\squared},
\SI{1e21}{\per\cm\squared},$ and
$\SI{1e20}{\per\cm\squared}$.
In addition, we vary the number of source iterations performed as $\nItScat = 2, 3, 4$ in order to check the convergence of the method.

We intuitively expect the shadow to become less and less prominent as the scattering cross section increases, due to the influx of scatter light on the low density gas.

The results of these tests at $t = \SI{40}{\kilo\year}$ are shown in \fref{scattering}.
Even for $\sigmaS = \SI{0}{\per\cm\squared}$, the ionized regions protrude substantially beyond the ideal shadow. This is the same numerical diffusion we already observed in Section~\sref{shadowing}. However, the shadow volume clearly decreases for increasing values of the scattering cross sections until the shadow vanishes almost entirely at $\sigmaS = \SI{5e21}{\per\cm\squared}$.
We also note that the number of iterations barely affects the result after $\nItScat = 2$, implying that the method converges rather quickly in this test case.

\begin{figure*}
    \includegraphics{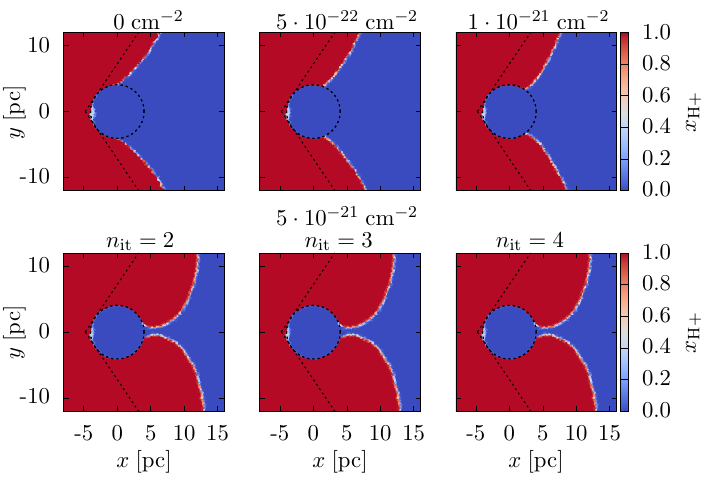}
    \caption{\label{fig:scattering}The abundance of ionized hydrogen $\xHP$ at $t = \SI{40}{\kilo\year}$ in a slice through the $z$-plane of the simulation box. The $x$ and $y$ axis show the $x$ and $y$ position in the box respectively. The dark dashed circle indicates the position of the over-dense clump. The dashed lines indicate indicate the boundaries of a hypothetical, perfectly sharp shadow.
      Top left: $\sigmaS = \SI{0}{\per\cm\squared}$,
      Top center: $\sigmaS = \SI{5e-22}{\per\cm\squared}$,
      Top right: $\sigmaS = \SI{1e-21}{\per\cm\squared}$,
      Bottom: $\sigmaS = \SI{5e21}{\per\cm\squared}$,
      Bottom left: $\nItScat = 2$,
      Bottom center: $\nItScat = 3$,
      Bottom right: $\nItScat = 4$,
    }
\end{figure*}

\subsection{Periodic Boundary Conditions\label{sec:pbcTest}}
As discussed in \secref{methodspbc}, Sweep handles periodic boundary conditions by an iterative scheme. In order to show that this scheme produces physical results, we perform a test similar to the R-type expansion in \secref{rtype}. We chose the case with a resolution of $n = 32^3$. The primary difference in this new test is the position of the point source, which we move to $\bm{r} = (6.336, 0, 0) \; \mathrm{kpc}$. Since the same box size of $L = \SI{12.8}{\kilo\parsec}$ is used, that corresponds to a source located very close to the right boundary of the simulation box. 

\begin{figure}
    \includegraphics{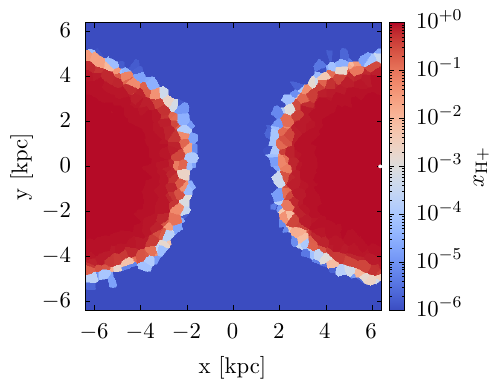}
    \caption{\label{fig:periodic_slice}A slice through the simulation box of the test described in \secref{pbcTest} at $z = 0$ and $t = \SI{58}{\mega\year} \approx 0.473 \cdot \trec$. The $x$ and $y$ axis show the $x$ and $y$ position in the box respectively. The color scale indicates the abundance of ionized hydrogen $\xHP$, with blue being neutral and red being fully ionized. The small white dot indicates the position of the source in the box.}
\end{figure}

\begin{figure}
    \includegraphics{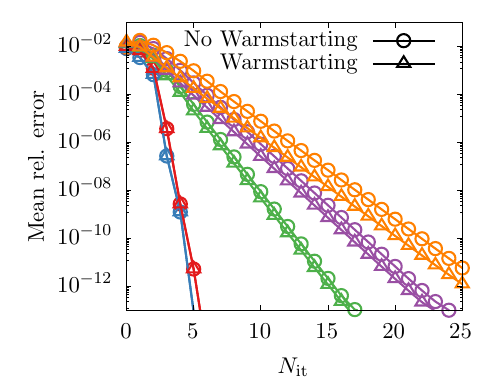}
    \caption{\label{fig:convergence_warmstarting}The mean relative error in the source terms (given by \eref{relativeErrorPbc}) as a function of the number of iterations $N_{\mathrm{it}}$ for the periodic boundary conditions test described in \secref{pbcTest}. The colors correspond to different time steps.
      Blue: $t = \SI{14.5}{\mega\year}$.
      Red: $t = \SI{29}{\mega\year}$.
      Green: $t = \SI{43.5}{\mega\year}$.
      Purple: $t = \SI{58}{\mega\year}$.
      Orange: $t = \SI{72.5}{\mega\year}$.
      Circles: Without warmstarting
      Triangles: With warmstarting
    }
\end{figure}

In \figref{convergence_warmstarting}, the mean relative error given by \eref{relativeErrorPbc} is shown as a function of the number of periodic boundary iterations at different times with and without warm starting.
The first clear trend that can be seen is that while the initial error remains roughly constant throughout time, the speed of the convergence decreases drastically.
While it takes $N_{\mathrm{it}} = 6$ iterations to reach an error of $E < 10^{-10}$ for the first timestep at $t = \SI{14.5}{\mega\year}$, it takes $N_{\mathrm{it}} = 14$ iterations to reach the same threshold at $t = \SI{43.5}{\mega\year}$.

We believe that this effect is partially due to re-entry dependencies - a cell very close the the right boundary at $x = \SI{6.4}{\kilo\parsec}$ will often have downwind dependencies at the left side of the boundary at $x = \SI{-6.4}{\kilo\parsec}$, especially for a sweep direction which is close to being contained within the $y$-$z$ plane. The cells on the other side of the boundary will then often have downwind neighbors on the right side of the boundary.
The effective distance traveled of photons along such re-entry dependencies is strongly determined by the number of iterations $N_{\mathrm{it}}$ since it takes one full iteration for the information about those photons to travel one cell.

This effect is exacerbated due to the location of the source in the test setup described above, since it is located very close to the boundary. This means that a high number of photons will be travelling along the boundary in a direction parallel to the $y$-$z$ plane.

In \figref{convergence_warmstarting}, the mean relative error given by \eref{relativeErrorPbc} is shown as a function of the number of periodic boundary iterations at different times with and without warmstarting.

As another test of the convergence of the iterative scheme, we calculate the radius of the ionized bubble as a function of time and compare the result to the analytical prediction.
The simulations in this test are equal to those in \secref{rtype}, with the only difference being the position of the source in the box, requiring the proper treatment of periodic boundary conditions in order to reproduce the behavior of the R-type expansion.
We choose the box with $32^3$ particles and perform simulations with iteration counts $1 \leq \nItPbc \leq 20$. All other parameters are chosen equal to those in \secref{rtype}.

\begin{figure}
    \includegraphics{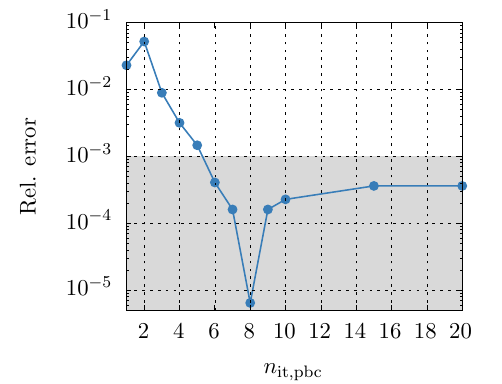}
    \caption{\label{fig:rtype_num_periodic_iterations}Relative error of the radius of the ionized sphere in the R-type expansion in a uniform medium with a source located at the boundary of the box as a function of the number of periodic boundary iterations $\nItPbc$. 
      The gray area signifies the approximate level of error expected due to the difference between numerical results and analytical prediction for a R-type expansion in the absence of periodic boundary conditions.
    }
\end{figure}

\figref{rtype_num_periodic_iterations} shows the relative error between the radius of the ionized sphere and the analytical prediction as a function of the number of periodic iterations.
As expected, the error decreases with the number of periodic iterations. After approximately 5 iterations, the error reaches values below $10^{-3}$, at which point it is indistinguishable from the error between the analytical prediction and the numerical results (see \fref{rtype}) which means that any discussion of the exact behavior of the error below that point is futile.

\subsection{Strong Scaling\label{sec:strongScaling}}
After the physical tests, we now discuss the scaling behavior of the Sweep method.
We begin by studying the strong scaling, i.e. the dependence of the time to solution $T$ of a problem of fixed time on the number $n$ of computing cores.
It is customary to study the scaling behavior of the code by comparing the time to solution $t(n)$ for a run on $n$ cores to the time to solution $\tbase$ for a base case at $\nbase$ (typically, $\nbase=1$) cores.
The time to solution of an ideally parallelized code $\tideal$ decreases as
\begin{equation}
  \label{eq:idealTime}
t_{\mathrm{ideal}}(n) = \frac{\tbase}{n / \nbase}.
\end{equation}

The parallel speedup $S$ is defined as
\begin{equation}
  S(n) = \frac{\tbase}{t(n)},
\end{equation}
and it follows from \eref{idealTime} that the speedup of an ideally parallelized code $\Sideal$ is given by
\begin{equation}
  \Sideal(n) = \frac{n}{\nbase}.
  \label{eq:strongScalingIdeal}
\end{equation}
We also define the parallel efficiency $\epsilon(n)$ as the fraction of the achieved speedup:
\begin{equation}
  \epsilon(n) = \frac{S(n)}{\Sideal} = \frac{S(n)}{n / \nbase}.
  \label{eq:parallelEfficiency}
\end{equation}

For these tests, we use the same simulation setup as in the shadowing test described in \secref{shadowing}. We study three different fixed problem sizes with $32^3$, $256^3$ and $512^3$ Voronoi cells respectively.
For each problem, we perform simulations for different numbers of cores. In the case of $32^3$ particles, we use a range from $n=1$ to $n=512$. For $256^3$ we use $n= 96$ to $n = 2048$ and for $512^3$, we perform runs from $n=2048$ to $n=8192$ cores. For the smallest case of $32^3$ particles, we compare our results to the SPRAI code.
We did not include a comparison to SPRAI for the larger problem sizes, since the run-time grew too large.

\begin{figure}
    \includegraphics{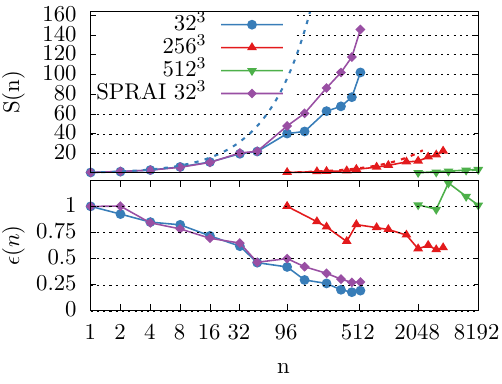}
    \caption{\label{fig:strongScalingSpeedup}
      Top: The parallel speedup of $S(n)$ as a function of the number of cores $n$. For Sweep, three problem sizes are shown: $32^3$ (blue circles), $256^3$ (red triangles), $512^3$ (green triangles). For SPRAI, we show the problem size $32^3$ (purple diamonds). For each problem size, the ideal, linear scaling behavior with respect to the base cases $\nbase=1$ for $32^3$, $\nbase=96$ for $256^3$ and $\nbase=2048$ for $512^3$ is given by \eref{strongScalingIdeal} is shown as the dashed line.
      Bottom: The parallel efficiency (defined in \eref{parallelEfficiency}) as a function of the number of cores for the same configurations. }
\end{figure}

In \fref{strongScalingSpeedup}, the parallel speedup is shown as a function of the number of compute cores in comparison to the ideal behavior given by \eref{strongScalingIdeal} for each of the three problem sizes. 
For low core numbers, both SPRAI and Sweep scale well with the number of cores. At $n=96$ cores the speedup of Sweep is $S(96) \approx 40$ with SPRAI being slightly faster at $S(96) \approx 48$, corresponding to parallel efficiencies of $\epsilon(96) = 42\%$ and $\epsilon(96) = 50\%$ respectively.
At higher core numbers, the rate of increase in the speedup declines, the parallel efficiency drops to $\epsilon(512) = 28\%$ for SPRAI and $20\%$ for Sweep.
This behavior is to be expected, since the ratio of the required inter-process communication to communicate the fluxes crossing processor domains to the amount of cells to solve locally decreases as the number of cores increases.

For the higher resolution runs, the rate of decrease in the efficiency of Sweep is lower.
Comparing the run with $256^3$ particles at $n=96$ to that with $n=2048$ shows a decrease in parallel efficiency to $\approx 60\%$.
For the run with $512^3$ particles, the parallel efficiency increases beyond $1$.
Such a result may initially seem counter-intuitive, but can be explained by the fact that for some numbers of cores the domain decomposition turns out to be particularly unfortunate, decreasing the efficiency of Sweep due to worse scheduling behavior or similar effects. If such a case is used as the reference simulation to which simulations at higher core numbers are compared the result are parallel efficiencies larger than $1$.

This, along with the fact that the real run-time of the code is hidden, highlights the fundamental problem with simply comparing the speedup of two codes without comparing their respective run times, since the parallel efficiency improves as the single-core performance of the parallel part of the code decreases. 

Therefore, it is important to show the run time of the code.
In order to compare the run times of different problem sizes in a reasonable manner, we define the time per task as
\begin{equation}
  \ttask(n) = \frac{n t(n)}{N_{\mathrm{dir}} N_{\mathrm{cells}} N_{\mathrm{freq}}}\;,
  \label{eq:timeTask}
\end{equation}
where $N_{\mathrm{dir}}$ is the number of directions ($84$ in our case), $N_{\mathrm{cells}}$ is the number of cells ($32^3$, $256^3$ and $512^3$, depending on the problem size) and $N_{\mathrm{freq}}$ is the number of frequencies. This is the effective time it takes a single core to solve a single cell in a single direction for a single frequency.
For an ideally parallelized code, $\ttask$ is independent of the number of cores.
This definition allows a comparison across different problem sizes by looking at the effective loss in performance given by $\ttask(1) / \ttask(n)$, which we believe is a realistic assessment of the performance of the code between runs of vastly different numbers of cores.

In \fref{strongScalingRuntime}, the time per task is shown as a function of the number of cores for the three different problem sizes. The figure shows that the two codes obtain very different run times on this particular test. Whereas the scaling behavior of the two codes are very similar, Sweep outperforms SPRAI by a factor of $\approx 20$ at $n=1$ cores.
It is important to note, however, that this result does not hold for any kind of simulation, since Sweep is written with a focus on simulations of reionization, where mean free path lengths are potentially high, while SPRAI performs comparatively well in dense media.

The lower panel of \fref{strongScalingRuntime} shows the effective performance $\ttask(1) / \ttask(n)$.
This demonstrates that, while the effective performance of Sweep decreases with the number of cores, it only decreases to $\approx 20\%$ at $n=8192$ cores.

\begin{figure}
    \includegraphics{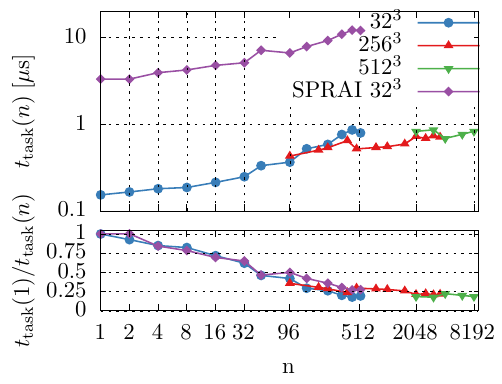}
    \caption{\label{fig:strongScalingRuntime}
     Top: The time per task $\ttask$ (see \eref{timeTask}) as a function of the number of cores $n$ for three problem sizes: $32^3$ (blue), $256^3$ (red), $512^3$ (green). 
     Bottom: The performance loss $\ttask(1) / \ttask(n)$ as a function of the number of cores for the same three problems. }
\end{figure}

\subsection{Weak Scaling\label{sec:weakScaling}}
As another test of the parallel efficiency of Sweep with increasing number of cores, we perform a weak scaling test by increasing the problem size in proportion to the number of cores, thus keeping the number of cells per core constant.
The speedup of an ideally parallelized algorithm in the weak scaling case is given by
\begin{equation}
  S(n) = 1.
  \label{eq:weakScalingIdeal}
\end{equation}

As a base case, we choose the $n=1$ case with a resolution of $32^3$ cells, identical to the corresponding $n=1$ simulation in the strong scaling test.
In addition to the base case we perform simulations for $n = 8, 48, 480$ ($528$ for SPRAI, due to memory requirements) and $4096$ cores with resolutions of $64^3$, $128^3$, $256^3$ and $512^3$ particles respectively\footnote{Note that the ratio of particles to cores does not remain exactly constant because of the number of cores was required to be divisible by 48}.
We do not include the case of $4096$ cores on $512^3$ particles for SPRAI, due to slightly increased memory requirements making a run on this number of cores difficult.

\begin{figure}
    \includegraphics{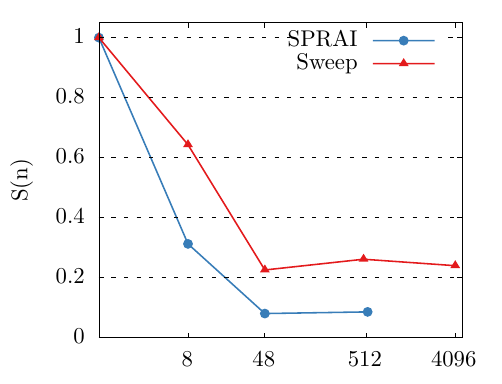}
    \caption{\label{fig:weakScalingSpeedup}The parallel speedup $S(n)$ as a function of the number of cores $n$ for a problem which scales in size with the number of cores for both Sweep (red) and SPRAI (blue). The ideal, linear scaling behavior given by \eref{weakScalingIdeal} is shown in green.}
\end{figure}

\figref{weakScalingSpeedup} shows the speedup as a function of the number of cores $n$, which, in the case of weak scaling is equivalent to the parallel efficiency. The speedup initially drops quite quickly, to values of $\sim \SI{22}{\percent}$ for Sweep and $\sim \SI{9}{\percent}$ for SPRAI at $n=48$ cores. However, the speedup does not decrease further and remains at similar values until $n=4096$ cores.

We believe that the initial decrease in efficiency is due to the overhead in communication compared to the base case of $n=1$ cores. In particular, the re-entry dependencies discussed in \secref{sweep} significantly slow down performance due to the amount of communication in which very few fluxes are exchanged.

While some parallel efficiency will be lost due to the increasing amount of communication for higher number of cores, another effect diminishing the parallel efficiency is the ``pipe fill'' described in \secref{sweep}, since cores whose domains lie in the inner regions of the simulation box cannot begin solving before those with domains in the outer regions have finished their sweep. At low numbers of cores ($n = 1$ or $n = 8$), no such domains exist, since every domain is adjacent to a boundary of the simulation box. As the number of cores increases, the number of inner regions increases and parallel efficiency decreases.
In order to check whether this effect is already affecting our results and decreasing the parallel efficiency significantly, we generated program output which displayed the timing at which the first task is solved for each core. This allowed us to estimate the amount of performance lost due to idle time. We found that for $n=4096$, this delay is still insignificant compared to the communication overhead.

\begin{figure}
    \includegraphics{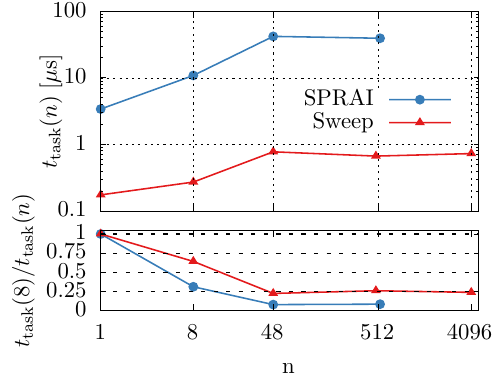}
    \caption{\label{fig:weakScalingRuntime}
     Top: The time per task $\ttask$ (see \eref{timeTask}) as a function of the number of cores $n$ in a weak scaling test where the problem size scales with the number of cores from $32^3$ at $n=1$ to $512^3$ at $n=4096$.
     Bottom: The performance loss $\ttask(1) / \ttask(n)$ as a function of the number of cores.
     Blue line: SPRAI, red line: Sweep.}
\end{figure}

\section{Conclusion}\label{sec:conclusion}
In this paper, we introduced a sweep-based radiative transfer method which we implemented for the moving-mesh hydrodynamics code Arepo. The method solves the radiative transfer equation under the assumption of an infinite speed of light and a steady state solution.
As a first test of our implementation of the code, we studied the expansion of an HII region around a point source for the R-type and the D-type regime and compared the results to the analytical predictions as well as results obtained with the SPRAI code.
We also performed a test which allowed us to study the shadowing behavior behind a dense blob of gas.
For all the tests, we find good agreement with our results and the results obtained via SPRAI.

In addition, we performed tests to better understand whether the source iteration method employed correctly deals with scattering.
To ensure that the code can also handle periodic boundary conditions we performed a series of tests similar to the R-type expansion but with a source very close to the boundaries of the box. We find that after $\sim$5 iterations of the periodic boundary sources, the results are virtually indistinguishable from those of the standard R-type test.

We also analyze the parallel efficiency of our code in order to assess whether large-scale simulations would be feasible with this method. To this end, we perform strong and weak scaling tests.
For the strong scaling, we find similar scaling behavior between SPRAI and Sweep for the smallest test case ($32^3$ particles), with Sweep outperforming SPRAI by a factor of ${\sim}10$ in the actual runtime.
We find a constant, slow decrease in the parallel efficiency down to ${\sim}20\%$ at $512$ cores for the smallest test case, however comparing the run time per cell between the large test cases ($512^3$ particles) at $8192$ cores and the smallest at $1$ cores, we find that Sweep still operates at ${\sim}20 \%$ efficiency.
In the case of weak scaling, the parallel efficiency decreases quite significantly with the number of cores. However, Sweep still performs better than SPRAI by a factor of $4$ at $n=512$ with a parallel efficiency of ${\sim}25 \%$ at $4096$ cores.
We expect the reduction in efficiency to be due to the fact that the domains of some cores are located in the inner region of the entire computational domain.

One possible measure to improve parallel efficiency is to change the domain decomposition. This is made more complicated due to the fact that sweep is intended to run in parallel to gravitation and hydrodynamics. In addition, estimating the amount of computational work for a single Voronoi cell is straightforward for gravitation and hydrodynamics, but difficult in general for radiative transfer, where assigning the cell to a certain core not only increases the total work load of that core but also affects the global scheduling problem.

Finally, we note that although we have developed Sweep with the goal of modeling cosmological reionization, the algorithm itself is far more general than this and could readily be adapted for use in other applications in which it is advantageous to have a method for modelling radiation transfer that is independent of the number of sources.

\section*{Acknowledgements}
We acknowledge computing resources and data storage facilities provided by the State of Baden-W\"{u}rttemberg through bwHPC and the German Research Foundation (DFG) through grant INST 35/1134-1 FUGG and INST 35/1503-1 FUGG. We also thank for computing time from the Leibniz Computing Center (LRZ) in project pr74nu. We thank for funding from the Heidelberg Cluster of Excellence EXC 2181 (Project-ID 390900948) `STRUCTURES: A unifying approach to emergent phenomena in the physical world, mathematics, and complex data' supported by the German Excellence Strategy. RSK, SCOG, and TP acknowledge financial support from the European Research Council in the ERC synergy grant `ECOGAL – Understanding our Galactic ecosystem: From the disk of the Milky Way to the formation sites of stars and planets' (project ID 855130),  from DFG via the Collaborative Research Center (SFB 881, Project-ID 138713538) 'The Milky Way System' (subprojects A1, B1, B2, B8), and from the German Ministry for Economic Affairs and Climate Action for funding in  project `MAINN -- Machine learning in Astronomy: understanding the physics of stellar birth with Invertible Neural Networks' (funding ID 50OO2206). We thank Joe Lewis, Dylan Nelson, and Annalisa Pillepich for useful discussions.

\section*{Data Availability}
The data underlying this article will be shared on reasonable request to the corresponding author.

\bibliographystyle{mnras}
\bibliography{ms}

\begin{thebibliography}{}
\makeatletter
\relax
\def\mn@urlcharsother{\let\do\@makeother \do\$\do\&\do\#\do\^\do\_\do\%\do\~}
\def\mn@doi{\begingroup\mn@urlcharsother \@ifnextchar [ {\mn@doi@}
  {\mn@doi@[]}}
\def\mn@doi@[#1]#2{\def\@tempa{#1}\ifx\@tempa\@empty \href
  {http://dx.doi.org/#2} {doi:#2}\else \href {http://dx.doi.org/#2} {#1}\fi
  \endgroup}
\def\mn@eprint#1#2{\mn@eprint@#1:#2::\@nil}
\def\mn@eprint@arXiv#1{\href {http://arxiv.org/abs/#1} {{\tt arXiv:#1}}}
\def\mn@eprint@dblp#1{\href {http://dblp.uni-trier.de/rec/bibtex/#1.xml}
  {dblp:#1}}
\def\mn@eprint@#1:#2:#3:#4\@nil{\def\@tempa {#1}\def\@tempb {#2}\def\@tempc
  {#3}\ifx \@tempc \@empty \let \@tempc \@tempb \let \@tempb \@tempa \fi \ifx
  \@tempb \@empty \def\@tempb {arXiv}\fi \@ifundefined
  {mn@eprint@\@tempb}{\@tempb:\@tempc}{\expandafter \expandafter \csname
  mn@eprint@\@tempb\endcsname \expandafter{\@tempc}}}

\bibitem[\protect\citeauthoryear{Abel, Norman  \& Madau}{Abel
  et~al.}{1999}]{abelPhotonconservingRadiativeTransfer1999}
Abel T.,  Norman M.~L.,   Madau P.,  1999, \mn@doi [The Astrophysical Journal]
  {10.1086/307739}, 523, 66

\bibitem[\protect\citeauthoryear{Adams et~al.,}{Adams
  et~al.}{2020}]{adamsProvablyOptimalParallel2020}
Adams M.~P.,  et~al., 2020, J. Comput. Phys., 407, 109234

\bibitem[\protect\citeauthoryear{Baczynski, Glover  \& Klessen}{Baczynski
  et~al.}{2015}]{baczynskiFerventChemistrycoupledIonizing2015}
Baczynski C.,  Glover S. C.~O.,   Klessen R.~S.,  2015, \mn@doi [Monthly
  Notices of the Royal Astronomical Society] {10.1093/mnras/stv1906}, 454, 380

\bibitem[\protect\citeauthoryear{Baker \& Koch}{Baker \&
  Koch}{1998}]{bakerSnAlgorithmMassively1998}
Baker R.~S.,  Koch K.~R.,  1998, \mn@doi [Nuclear Science and Engineering]
  {10.13182/NSE98-1}, 128, 312

\bibitem[\protect\citeauthoryear{{Boss}}{{Boss}}{2008}]{Boss2008}
{Boss} A.~P.,  2008, \mn@doi [\apj] {10.1086/533496}, \href
  {https://ui.adsabs.harvard.edu/abs/2008ApJ...677..607B} {677, 607}

\bibitem[\protect\citeauthoryear{Chang, Davis  \& Jiang}{Chang
  et~al.}{2020}]{changTimedependentRadiationHydrodynamics2020}
Chang P.,  Davis S.~W.,   Jiang Y.-F.,  2020, \mn@doi [Monthly Notices of the
  Royal Astronomical Society] {10.1093/mnras/staa573}, 493, 5397

\bibitem[\protect\citeauthoryear{Dullemond, Juhasz, Pohl, Sereshti, Shetty,
  Peters, Commercon  \& Flock}{Dullemond
  et~al.}{2012}]{dullemondRADMC3DMultipurposeRadiative2012}
Dullemond C.~P.,  Juhasz A.,  Pohl A.,  Sereshti F.,  Shetty R.,  Peters T.,
  Commercon B.,   Flock M.,  2012, Astrophysics Source Code Library, p.
  ascl:1202.015

\bibitem[\protect\citeauthoryear{{Gnedin} \& {Abel}}{{Gnedin} \&
  {Abel}}{2001}]{Gnedin01}
{Gnedin} N.~Y.,  {Abel} T.,  2001, \mn@doi [\na]
  {10.1016/S1384-1076(01)00068-9}, \href
  {https://ui.adsabs.harvard.edu/abs/2001NewA....6..437G} {6, 437}

\bibitem[\protect\citeauthoryear{Hartwig, Glover, Klessen, Latif  \&
  Volonteri}{Hartwig et~al.}{2015}]{hartwigHowImprovedImplementation2015}
Hartwig T.,  Glover S. C.~O.,  Klessen R.~S.,  Latif M.~A.,   Volonteri M.,
  2015, \mn@doi [Monthly Notices of the Royal Astronomical Society]
  {10.1093/mnras/stv1368}, 452, 1233

\bibitem[\protect\citeauthoryear{{Hayes} \& {Norman}}{{Hayes} \&
  {Norman}}{2003}]{Hayes2003}
{Hayes} J.~C.,  {Norman} M.~L.,  2003, \mn@doi [\apjs] {10.1086/374658}, \href
  {https://ui.adsabs.harvard.edu/abs/2003ApJS..147..197H} {147, 197}

\bibitem[\protect\citeauthoryear{{Iliev}, {Mellema}, {Ahn}, {Shapiro}, {Mao}
  \& {Pen}}{{Iliev} et~al.}{2014}]{iliev14}
{Iliev} I.~T.,  {Mellema} G.,  {Ahn} K.,  {Shapiro} P.~R.,  {Mao} Y.,   {Pen}
  U.-L.,  2014, \mn@doi [Monthly Notices of the Royal Astronomical Society]
  {10.1093/mnras/stt2497}, \href
  {https://ui.adsabs.harvard.edu/abs/2014MNRAS.439..725I} {439, 725}

\bibitem[\protect\citeauthoryear{Jaura, Glover, Klessen  \& Paardekooper}{Jaura
  et~al.}{2018}]{jauraSPRAICouplingRadiative2018}
Jaura O.,  Glover S. C.~O.,  Klessen R.~S.,   Paardekooper J.-P.,  2018,
  \mn@doi [Monthly Notices of the Royal Astronomical Society]
  {10.1093/mnras/stx3356}, 475, 2822

\bibitem[\protect\citeauthoryear{Jaura, Magg, Glover  \& Klessen}{Jaura
  et~al.}{2020}]{jauraSPRAIIIMultifrequencyRadiative2020}
Jaura O.,  Magg M.,  Glover S. C.~O.,   Klessen R.~S.,  2020, \mn@doi [Monthly
  Notices of the Royal Astronomical Society] {10.1093/mnras/staa3054}, 499,
  3594

\bibitem[\protect\citeauthoryear{Jiang, Stone  \& Davis}{Jiang
  et~al.}{2014}]{jiangAlgorithmRadiationMagnetohydrodynamics2014}
Jiang Y.-F.,  Stone J.~M.,   Davis S.~W.,  2014, \mn@doi [The Astrophysical
  Journal Supplement Series] {10.1088/0067-0049/213/1/7}, 213, 7

\bibitem[\protect\citeauthoryear{Kannan, Vogelsberger, Marinacci, McKinnon,
  Pakmor  \& Springel}{Kannan
  et~al.}{2019}]{kannanArepoRTRadiationHydrodynamics2019}
Kannan R.,  Vogelsberger M.,  Marinacci F.,  McKinnon R.,  Pakmor R.,
  Springel V.,  2019, \mn@doi [Monthly Notices of the Royal Astronomical
  Society] {10.1093/mnras/stz287}, 485, 117

\bibitem[\protect\citeauthoryear{{Kim}, {Kim}  \& {Ostriker}}{{Kim}
  et~al.}{2018}]{kim18}
{Kim} J.-G.,  {Kim} W.-T.,   {Ostriker} E.~C.,  2018, \mn@doi [The
  Astrophysical Journal] {10.3847/1538-4357/aabe27}, \href
  {https://ui.adsabs.harvard.edu/abs/2018ApJ...859...68K} {859, 68}

\bibitem[\protect\citeauthoryear{Koch, Baker  \& Alcouffe}{Koch
  et~al.}{1991}]{kochSolutionFirstorderForm1991}
Koch K.~R.,  Baker R.~S.,   Alcouffe R.~E.,  1991, Transactions of the American
  Nuclear Society, 65, 198

\bibitem[\protect\citeauthoryear{Kruip, Paardekooper, Clauwens  \& Icke}{Kruip
  et~al.}{2010}]{kruipMathematicalPropertiesSimpleX2010}
Kruip C. J.~H.,  Paardekooper J.-P.,  Clauwens B. J.~F.,   Icke V.,  2010,
  \mn@doi [Astronomy and Astrophysics] {10.1051/0004-6361/200913439}, 515, A78

\bibitem[\protect\citeauthoryear{{Krumholz}, {Klein}  \& {McKee}}{{Krumholz}
  et~al.}{2007}]{Krumholz2007}
{Krumholz} M.~R.,  {Klein} R.~I.,   {McKee} C.~F.,  2007, \mn@doi [\apj]
  {10.1086/510664}, \href
  {https://ui.adsabs.harvard.edu/abs/2007ApJ...656..959K} {656, 959}

\bibitem[\protect\citeauthoryear{Levermore \& Pomraning}{Levermore \&
  Pomraning}{1981}]{levermoreFluxlimitedDiffusionTheory1981}
Levermore C.~D.,  Pomraning G.~C.,  1981, \mn@doi [The Astrophysical Journal]
  {10.1086/159157}, 248, 321

\bibitem[\protect\citeauthoryear{Loeb \& Barkana}{Loeb \&
  Barkana}{2001}]{loebReionizationUniverseFirst2001}
Loeb A.,  Barkana R.,  2001, \mn@doi [Annual Review of Astronomy and
  Astrophysics] {10.1146/annurev.astro.39.1.19}, 39, 19

\bibitem[\protect\citeauthoryear{Lucero~Lorca}{Lucero~Lorca}{2018}]{lucerolorcaMultilevelSchwarzMethods2018}
Lucero~Lorca J.~P.,  2018, PhD thesis, Heidelberg University

\bibitem[\protect\citeauthoryear{Mihalas \& {Weibel-Mihalas}}{Mihalas \&
  {Weibel-Mihalas}}{1999}]{mihalasFoundationsRadiationHydrodynamics1999}
Mihalas D.,  {Weibel-Mihalas} B.,  1999, Foundations of Radiation
  Hydrodynamics.
{Dover}, {Mineola, N.Y}

\bibitem[\protect\citeauthoryear{Oxley \& Woolfson}{Oxley \&
  Woolfson}{2003}]{oxleySmoothedParticleHydrodynamics2003}
Oxley S.,  Woolfson M.~M.,  2003, \mn@doi [Monthly Notices of the Royal
  Astronomical Society] {10.1046/j.1365-8711.2003.06751.x}, 343, 900

\bibitem[\protect\citeauthoryear{Paardekooper, Kruip  \& Icke}{Paardekooper
  et~al.}{2010}]{paardekooperSimpleX2RadiativeTransfer2010}
Paardekooper J.-P.,  Kruip C. J.~H.,   Icke V.,  2010, \mn@doi [Astronomy and
  Astrophysics] {10.1051/0004-6361/200913821}, 515, A79

\bibitem[\protect\citeauthoryear{Pautz}{Pautz}{2002}]{pautzAlgorithmParallelSweeps2002}
Pautz S.~D.,  2002, \mn@doi [Nuclear Science and Engineering]
  {10.13182/NSE02-1}, 140, 111

\bibitem[\protect\citeauthoryear{{Peters}, {Banerjee}, {Klessen}, {Mac Low},
  {Galv{\'a}n-Madrid}  \& {Keto}}{{Peters} et~al.}{2010}]{peters10}
{Peters} T.,  {Banerjee} R.,  {Klessen} R.~S.,  {Mac Low} M.-M.,
  {Galv{\'a}n-Madrid} R.,   {Keto} E.~R.,  2010, \mn@doi [The Astrophysical
  Journal] {10.1088/0004-637X/711/2/1017}, \href
  {https://ui.adsabs.harvard.edu/abs/2010ApJ...711.1017P} {711, 1017}

\bibitem[\protect\citeauthoryear{Ritzerveld \& Icke}{Ritzerveld \&
  Icke}{2006}]{ritzerveldTransportAdaptiveRandom2006}
Ritzerveld J.,  Icke V.,  2006, \mn@doi [Physical Review E]
  {10.1103/PhysRevE.74.026704}, 74, 026704

\bibitem[\protect\citeauthoryear{Rosdahl, Blaizot, Aubert, Stranex  \&
  Teyssier}{Rosdahl et~al.}{2013}]{rosdahlRamsesrtRadiationHydrodynamics2013}
Rosdahl J.,  Blaizot J.,  Aubert D.,  Stranex T.,   Teyssier R.,  2013, \mn@doi
  [Monthly Notices of the Royal Astronomical Society] {10.1093/mnras/stt1722},
  436, 2188

\bibitem[\protect\citeauthoryear{Rybicki \& Lightman}{Rybicki \&
  Lightman}{1985}]{rybickiRadiativeProcessesAstrophysics1985}
Rybicki G.~B.,  Lightman A.~P.,  1985, Radiative {{Processes}} in
  {{Astrophysics}}.
{Wiley-VCH Verlag GmbH \& Co. KGaA}, {Weinheim, Germany},
  \mn@doi{10.1002/9783527618170}

\bibitem[\protect\citeauthoryear{Saad \& Schultz}{Saad \&
  Schultz}{1986}]{saadGMRESGeneralizedMinimal1986}
Saad Y.,  Schultz M.~H.,  1986, \mn@doi [SIAM Journal on Scientific and
  Statistical Computing] {10.1137/0907058}, 7, 856

\bibitem[\protect\citeauthoryear{Schauer, Glover, Klessen  \& Ceverino}{Schauer
  et~al.}{2019}]{schauerInfluenceStreamingVelocities2019}
Schauer A. T.~P.,  Glover S. C.~O.,  Klessen R.~S.,   Ceverino D.,  2019,
  \mn@doi [Monthly Notices of the Royal Astronomical Society]
  {10.1093/mnras/stz013}, 484, 3510

\bibitem[\protect\citeauthoryear{Smith, Kannan, Tsang, Vogelsberger  \&
  Pakmor}{Smith et~al.}{2020}]{smithAREPOMCRTMonteCarlo2020}
Smith A.,  Kannan R.,  Tsang B. T.~H.,  Vogelsberger M.,   Pakmor R.,  2020,
  \mn@doi [The Astrophysical Journal] {10.3847/1538-4357/abc47e}, 905, 27

\bibitem[\protect\citeauthoryear{Spitzer}{Spitzer}{1978}]{spitzerPhysicalProcessesInterstellar1978}
Spitzer L.,  1978, Physical {{Processes}} in the {{Interstellar Medium}}, first
  edn.
{Wiley}, \mn@doi{10.1002/9783527617722}

\bibitem[\protect\citeauthoryear{Springel}{Springel}{2010}]{springelPurSiMuove2010}
Springel V.,  2010, \mn@doi [Monthly Notices of the Royal Astronomical Society]
  {10.1111/j.1365-2966.2009.15715.x}, 401, 791

\bibitem[\protect\citeauthoryear{Str{\"o}mgren}{Str{\"o}mgren}{1939}]{stromgrenPhysicalStateInterstellar1939}
Str{\"o}mgren B.,  1939, \mn@doi [The Astrophysical Journal] {10.1086/144074},
  89, 526

\bibitem[\protect\citeauthoryear{Vermaak, Ragusa, Adams  \& Morel}{Vermaak
  et~al.}{2021}]{vermaakMassivelyParallelTransport2020}
Vermaak J. I.~C.,  Ragusa J.~C.,  Adams M.~L.,   Morel J.~E.,  2021, J. Comput.
  Phys., 425, 109892

\bibitem[\protect\citeauthoryear{Whalen \& Norman}{Whalen \&
  Norman}{2006}]{whalenMultistepAlgorithmRadiation2006}
Whalen D.,  Norman M.~L.,  2006, \mn@doi [The Astrophysical Journal Supplement
  Series] {10.1086/499072}, 162, 281

\bibitem[\protect\citeauthoryear{{Whitehouse} \& {Bate}}{{Whitehouse} \&
  {Bate}}{2004}]{whitehouseSmoothedParticleHydrodynamics2004}
{Whitehouse} S.~C.,  {Bate} M.~R.,  2004, \mn@doi [\mnras]
  {10.1111/j.1365-2966.2004.08131.x}, \href
  {https://ui.adsabs.harvard.edu/abs/2004MNRAS.353.1078W} {353, 1078}

\bibitem[\protect\citeauthoryear{{Wise}}{{Wise}}{2019}]{wise19}
{Wise} J.~H.,  2019, \mn@doi [Contemporary Physics]
  {10.1080/00107514.2019.1631548}, \href
  {https://ui.adsabs.harvard.edu/abs/2019ConPh..60..145W} {60, 145}

\bibitem[\protect\citeauthoryear{{Zaroubi}}{{Zaroubi}}{2013}]{Zaroubi13}
{Zaroubi} S.,  2013, in {Wiklind} T.,  {Mobasher} B.,   {Bromm} V.,  eds,
  Astrophysics and Space Science Library Vol. 396, The First Galaxies. p.~45

\bibitem[\protect\citeauthoryear{Zeyao \& Lianxiang}{Zeyao \&
  Lianxiang}{2004}]{zeyaoParallelFluxSweep2004}
Zeyao M.,  Lianxiang F.,  2004, \mn@doi [The Journal of Supercomputing]
  {10.1023/B:SUPE.0000032778.36178.d8}, 30, 5

\makeatother
\end{thebibliography}

\appendix
\section{Proof that sweep dependency graphs induced by Voronoi grids are acyclic}
\label{sec:voronoiCycleProof}
The solution of a cell $c^{\prime}$ in a sweep in the direction $\direc$ depends on fluxes of a neighboring cell $c$ if the normal $n$ of the face connecting $c$ and $c^{\prime}$ (defined such that it points towards $c^{\prime}$) fulfills
\begin{align}
  \bm{n} \cdot \direc > 0. \label{eq:downwindDefinition}
\end{align}
In a Voronoi grid, the face normal $\bm{n}$ is given by
\begin{align}
  \bm{n} = \frac{\bm{p}^{\prime} - \bm{p}}{\abs{\bm{p}^{\prime} - \bm{p}}}, \label{eq:normalDefinition}
\end{align}
where $p$ and $p^{\prime}$ are the Delaunay points corresponding to the respective Voronoi cells.

Now assume that there is a cycle $c_1, c_2, \dots, c_n, c_{n+1}$ in the dependencies, such that each cell in the cycle depends on the next and $c_{n+1}$ is the same cell as $c_1$.
Now, clearly
\begin{align}
  \bm{\Omega} \cdot \bb{\sum_{i=1}^{n} \bb{\bm{p}_{i+1} - \bm{p}_{i}}} = \bm{\Omega} \cdot \bm{0},
  \label{eq:contradiction}
\end{align}
but combining \eref{downwindDefinition} with \eref{normalDefinition} yields that each term in the sum in \eref{contradiction} is larger than zero, which is a contradiction.
Therefore, there are no cycles in the sweep dependencies induced by a Voronoi grid.

\bsp	
\label{lastpage}
\end{document}